\documentclass[a4paper,12pt]{article}  
\pdfoutput=1
\usepackage{hyperref}
\usepackage{epsfig}
\usepackage{amssymb}
\usepackage{cancel}
\usepackage{setspace}
\usepackage[usenames,dvipsnames,svgnames,table]{xcolor}
\usepackage{amsmath}
\usepackage{graphicx}
\usepackage{subfigure}
\usepackage{url}
\usepackage{slashed}
\usepackage{booktabs}
\usepackage{graphbox}
\usepackage{placeins}

\newcommand{\be}{\begin{equation}}
\newcommand{\ee}{\end{equation}}
\newcommand{\br}{\begin{eqnarray}}
\newcommand{\bea}{\begin{eqnarray}}
\newcommand{\eea}{\end{eqnarray}}
\newcommand{\er}{\end{eqnarray}}
\newcommand{\ba}{\begin{array}}
\newcommand{\ea}{\end{array}}
\newcommand{\bi}{\begin{itemize}}
\newcommand{\ei}{\end{itemize}}
\newcommand{\bn}{\begin{enumerate}}
\newcommand{\en}{\end{enumerate}}
\newcommand{\bc}{\begin{center}}
\newcommand{\ec}{\end{center}}

\newcommand{\beq}{\begin{equation}}
\newcommand{\eeq}{\end{equation}}

\newcommand{\U}{\scriptscriptstyle U}
\newcommand{\D}{\scriptscriptstyle D}

\newcommand{\gsim}{\lower1.0ex\hbox{$\;\stackrel{\textstyle>}{\sim}\;$}}
\newcommand{\lsim}{\lower1.0ex\hbox{$\;\stackrel{\textstyle<}{\sim}\;$}}

\newcommand{\bs}{\begin{small}}
\newcommand{\es}{\end{small}}

\newcommand{\qui}{q_{{\scriptscriptstyle U}_{\!i}}}
\newcommand{\qdi}{q_{{\scriptscriptstyle D}_{\!i}}}

\newcommand{\dg}{\dagger}

\newcommand{\g}{g_{\scriptscriptstyle{LR}}}

\begin{document}
\begin{center}
{\Large {\bf Dark origin of quark flavor hierarchy and mixing
}}
\\
\vspace*{1.5cm}
{
\large     

 { Emidio Gabrielli$^{a,b,c}$},  { Carlo Marzo$^{c}$}, { Luca Marzola$^{c}$}   and { Kristjan M\"u\"ursepp$^{d}$}
}\\
 \vspace{0.5cm}
 {\it
 (a)  Dipartimento di Fisica, Theoretical section, Universit\`a di 
 Trieste, \\ Strada Costiera 11, I-34151 Trieste, Italy \\ 
 (b) INFN, Sezione di Trieste, Via Valerio 2, I-34127 Trieste, Italy
 \\
 (c) NICPB, R\"avala 10, Tallinn 10143, Estonia 
 \\
 (d) Institute of Physics,
 University of Tartu, W. Ostwaldi 1, 50411 Tartu, Estonia.%
\\[1mm] }
\vspace*{2cm}{\bf ABSTRACT}
\end{center}
\vspace{0.3cm}

\noindent
We provide a dynamical mechanism for the generation of the Cabibbo-Kobayashi-Maskawa matrix in the context of a recently proposed model of flavor. The framework, based on the paradigm that Yukawa couplings are effective low energy couplings generated radiatively by dark sector interactions, is here extended to include a new scalar field which plays the role of a {\it dark flavon}.
Being singlet under the Standard Model gauge group but charged under the $U(1)$ symmetry of the dark sector, this particle sources new two- and three-loop diagrams that result in off-diagonal Yukawa interactions, thereby providing a simple explanation for the observed structure of the CKM matrix. By using an effective parametrization of these new loop contributions, we show that the CKM matrix elements can be correctly reproduced for perturbative values of the involved couplings. We outline the rich phenomenology predicted by the scenario and detail its implications for the LHC and future collider experiments.

\newpage

\section{Introduction}
\label{sec:Intro}
The origin of flavor hierarchy, reflected in the elementary fermion mass spectrum, remains one of the big conundrums of Nature. The precise LHC measurements of the Higgs boson couplings to gauge bosons and third generation fermions strongly support the Standard Model (SM) mechanism of fermion mass generation, based on the Higgs Yukawa couplings and the electroweak symmetry breaking (EWSM) \cite{Aad:2012tfa,ATLAS:2019slw,Sirunyan:2018koj}. The SM, however, does not shine any light on the puzzle related to the observed span of Yukawa couplings, should this be related to an unknown underlying mechanism rather than the outcome of free parameters not regulated by any symmetry. On top of that, the structure of the Cabibbo-Kobayashi-Maskawa (CKM) matrix measured in the quark weak interactions \cite{Tanabashi:2018oca} further complicates the overall picture and seems to call for a fundamental mechanism to implement, at least in part, the minimal Flavor violation ansatz. 
  
The large hierarchy in the fermion masses, as well as the absence of any underlying symmetry in the Yukawa sector, strongly suggest that these interactions could be effective low energy couplings rather then fundamental parameters. 
The first attempt in this direction dates back to the Froggat-Nielsen mechanism \cite{Froggatt:1978nt}, which relies on higher dimensional operators involving a progressive number of scalar flavon fields $\phi$ charged under a new $U(1)_F$ symmetry. In this scheme, the low-energy SM Yukawa couplings then scale as powers of $(\langle\phi\rangle/\Lambda)^n$, where $<\phi>$ is the vacuum expectation value (vev) of the corresponding flavon field, $\Lambda$ is the effective cut-off scale, and the exponent $n$ depends on the $U(1)_F$ charges of the involved fields. It is presently not clear which concrete models of new physics could implement such a construction, and, given the high dimensionality of the operators required to reproduce the full SM fermion spectrum and mixing, whether the mechanism itself could be eventually tested.
 
A more recent alternative put forward a different solution to the flavor hierarchy puzzle based on the generation of exponentially spread Yukawa couplings \cite{Gabrielli:2013jka,Gabrielli:2016vbb}. In this scenario, the SM is extended to include a dark sector consisting of a set of dark fermions singlet under the SM gauge interactions. New heavy scalar messenger fields, which instead carry the same internal quantum numbers as quarks and leptons, provide generation-blind interactions that bridge the dark and SM sectors. In this construction, the messengers and the dark fermions are charged under a {\it dark} $U(1)_{\D}$ gauge symmetry, accompanied by the corresponding massless gauge photon: the \textit{dark-photon}. Phenomenological implications of this scenario have been analyzed in the context of dark photon searches ~\cite{Biswas:2017lyg,Biswas:2016jsh,Biswas:2015sha,Gabrielli:2014oya} and $Z$ boson physics~\cite{Fabbrichesi:2017zsc}. 

In more detail, the SM Yukawa couplings are enforced to vanish at the tree-level by using a new symmetry $S$, for instance $SU(2)_L\times  SU(2)_R$ in~\cite{Gabrielli:2016vbb}. The interactions in the mediator sectors then induce finite loop contributions that result, after the spontaneous symmetry breaking of the $S$ symmetry, in effective low-energy Yukawa couplings $Y_i$ proportional to the corresponding dark fermion masses $M_i$. Explicitly, we have $Y_i\sim M_i/\Lambda_{\rm eff}$, where $\Lambda_{\rm eff}$ is an almost universal effective scale. Crucially, a non-perturbative $U(1)_D$ dynamics~\cite{Gabrielli:2007cp} in the dark sector generates the desired exponential spread in dark fermion spectrum, leading in the weak $\alpha_D$ coupling regime to \cite{Gabrielli:2007cp,Gabrielli:2013jka,Gabrielli:2016vbb}
\bea
Y_i\sim M_i\sim \exp{\left(-\frac{\gamma}{q^2_i\alpha_D}\right)}\,.
\eea
The fundamental parameters $q_i$ in the argument of the exponential are related to the $U(1)_D$ dark fermion charges, with $\alpha_D$ being the corresponding fine structure constant. The parameter $\gamma$ is instead related to an anomalous dimension. As a consequence, charges $q_i$ of the order
${\cal O}(1)$ can easily fit the SM spectrum, possibly including Dirac neutrino masses, and naturally explain the origin of the SM Flavor hierarchy. Alternatively, a similar exponential scaling law can be achieved in the strongly coupled $U(1)_D$ regime via the Miransky mechanism for chiral symmetry breaking \cite{Miransky:1984ef,Gabrielli:2016vbb}.

In order to also generate the CKM mixing, the portal interaction between the generic couplings of SM fermions $q^i_{L/R}$, dark fermions $Q^j$, and messengers fields $S^j$ must have the form \cite{Gabrielli:2016cut}
\bea
    {\cal L} \sim \hat{g}_L \bar{q}^i_L X^L_{ij} Q_R^{j}\hat{S}^{j}_{L} + \left\{L\leftrightarrow R\right\}\, ,
\label{portal}
\eea
where $\hat{g}_{L,R}$ are the corresponding couplings, $X_{ij}$ is a generic matrix and the sum over flavor ($i,j$) as well as other internal indices (spin, color) is understood. The misalignment between flavour and mass eigenstates that sources the CKM mixing is then ascribed to the $X$ matrix and the diagonal terms in the Yukawa matrices can be recovered by simply setting $X_{ij}\to \delta_{ij}$.

On general grounds, the off-diagonal contributions into the $X_{ij}$ terms arise after a misalignment between the mass matrices of dark fermions and quark fields. This observation has been used to investigate new processes related to flavor changing neutral currents~\cite{Gabrielli:2016cut} and kaon physics~\cite{Fabbrichesi:2017vma,Barducci:2018rlx}.  
In order to reproduce the observed hierarchy in the CKM matrix, it is sufficient that the structure of the $X_{ij}$ matrix obey the minimal flavor violation (MFV) hypothesis: $X_{ij}=\delta_{ij}+\Delta_{ij}$, with the off-diagonal terms $\Delta_{ij}\ll 1$ for $i\neq j$. Indeed, although consistent with observations, this ansatz does not motivate the $\Delta_{ij}\ll 1$ condition and additional {\it ad-hoc} constraints must, therefore, be imposed \cite{Gabrielli:2016vbb}.

The main objective of the present paper is to overcome this problem, exploring a new mechanism that dynamically generates off-diagonal terms in the Yukawa couplings which naturally respects the MFV condition and recovers the observed CKM matrix structure. Our idea is that the off-diagonal terms in the Yukawa matrix must be generated at higher orders in perturbation theory with respect to the diagonal entries, which are still the result of one-loop processes.

To this purpose, we restrict the interaction in Eq.(\ref{portal}) to universal diagonal couplings by setting $X_{ij}\to \delta_{ij}$, and then generate small off-diagonal contributions via higher order radiative processes. The mechanism is implemented by a set of scalar fields singlets under the SM gauge group, the {\it dark flavons}, that mediate $U(1)_D$ charged transitions between the dark fermions or the messenger fields. The off-diagonal terms modelled in $X_{ij}$ then arise via the exchange of dark flavons inside the one-loop diagrams used to generate the diagonal elements. In particular, the off-diagonal terms that connect the first two SM generation of up (or down) quarks, $Y_{12}$, as well as the one between the second and third generation, $Y_{23}$, arise at the two-loop level. The remaining $Y_{13}$, coupling, that connects the first and third generation, arises instead only at the 3-loop level. In this way, the obtained loop structure induces an hierarchy between the Yukawa matrix entries, which in turn results in the observed CKM matrix structure.

As a first check of the scenario, we propose an effective parametrization of the obtained Yukawa matrix texture and constrain it by letting the parameters in the CKM matrix to vary in their observed ranges. We then derive analytical expressions for the same Yukawa entries by evaluating the underlying two and three-loop Feynman diagrams induced by the dark flavon insertions. Finally, by matching these theoretical predictions against the results obtained with the effective parametrization, we detail how the observed quark mixing and flavour hierarchy bound the parameter space of the model. We also outline possible experimental tests of the scenario, which fall well within the discovery range of the present experiments at the LHC and future colliders. We anticipate that the main difference between the phenomenology of this extended model and that of the original framework proposed in~\cite{Gabrielli:2016vbb,Gabrielli:2016cut}, concerns the stability of the dark fermions. In particular, in the original scenario, the dark fermions decay is forbidden by the $U(1)_D$ gauge invariance because of the absence of $U(1)_D$ charged currents. On the contrary, in the model at hand, the presence of dark flavons allows the heaviest of dark fermions to decay and, as a consequence, only the lightest of these states can be a potential candidates for dark matter.

The paper is organized as follows: in Section 2 we summarize the features of the model and detail the novelties of the scenario analyzed. In Section 3 we introduce the effective parametrization of the obtained Yukawa coupling matrices, as well as the corresponding analytical expressions, and show how flavour experiments constrain the framework. In Section 4 we briefly discuss the new phenomenological aspects of the model. Our conclusions are presented in Section 5.

\section{Theoretical framework} 
\label{sec:Radiative Yukawa couplings}
We summarize here the main features of the model at the basis of the present work, originally proposed in \cite{Gabrielli:2013jka}, or in \cite{Gabrielli:2016vbb} within the context of Left-Right (LR) gauge symmetry.

As mentioned in the Introduction, the radiative generation of Yukawa couplings requires the presence of a hidden or dark sector external to the SM. This new sector must provide the necessary chiral symmetry breaking and contain a set of (heavy) messenger fields that connect with the SM fields.
Within the present framework, the dark sector consists of massive Dirac fermions, singlets under the SM gauge interactions (or, in the extended left-right (LR) version, singlet under the $SU(2)_L \times SU(2)_R\times U(1)_Y$ gauge group), but charged under a new, unbroken, $U(1)_D$ dark interaction. The presence of an unbroken $U(1)_D$ is necessary in order to generate the required exponential spread of dark fermion masses via non-perturbative dynamics, see \cite{Gabrielli:2007cp,Gabrielli:2014oya} for more details, and the construction therefore establishes a one-to one correspondence between the dark fermions and SM fermions.

We use here the LR gauge symmetry group in order to forbid the emergence of SM Yukawa operators at tree-level, although this can also be achieved by simply imposing a $Z_2$ discrete symmetry as originally proposed in \cite{Gabrielli:2013jka}. Consequently, Yukawa couplings can only arise as low energy effective operators after the symmetry that forbids the SM Yukawa couplings ($SU(2)_R$ in the present case) is spontaneously broken. The source of the chiral symmetry breaking is here provided by the dark fermion masses, that we take as free parameters of the theory. For the sake of simplicity we now restrict our discussion to the quark sector, remarking however that the extension to the lepton sector is straightforward.

Due to the quark SM quantum numbers, the minimal matter content needed for the colored messenger scalar sector is uniquely predicted, and for the $SU(2)_L\times SU(2)_R\times U(1)_Y$ case it is given by

\begin{itemize}
\item $2N$ complex scalar $SU(2)_L$ doublets: $\hat{S}_L^{\U_i}$ and $\hat{S}_L^{\D_i}$,
\item $2N$ complex scalar  $SU(2)_R$ doublets: $\hat{S}_R^{\U_i}$ and $\hat{S}_R^{\D_i}$,
\end{itemize} 
where
$\hat{S}_{A}^{\U_i,\D_i}=\left(\begin{array}{c}S^{\U_i,\D_i}_{A,1}\\S^{\U_i,\D_i}_{A,2}
\end{array}\right)$,  with $A=\{L,R\}$, 
$N=3$ and $i=1,2,3$ indicating the generation. The $L,R$ labels identify the messenger fields which couple to the SM fermions of that chirality, in complete analogy with the nomenclature of squark fields within supersymmetric theories.
The $\hat{S}_{L,R}^{\U_i,\D_i}$ fields then carry the same quantum numbers as the SM quarks of chirality $L,R$, and interact with the electroweak gauge bosons and gluons via their covariant derivatives. We summarize in Table~\ref{tab1} the relevant quantum numbers of the dark fermions and messenger scalar fields.

In order to maximize the symmetries of the theory, we require the Lagrangian to be invariant under a global $SU(N_F)$, where $N_F$ is the number of considered flavors. The messenger Lagrangian will then comprise only 4 different  
universal mass terms in both the $\hat{S}_{L,R}^{\U_i}$ and $\hat{S}_{L,R}^{\D_i}$ sectors. The symmetry can be further extended to accommodate parity and enforce a common scalar mass scale in the two sectors without raising phenomenological issues.

\begin{table} \begin{center}    
\begin{tabular}{ccccccc}
\toprule 
Fields 
& Spin
& $SU(2)_L$ 
& $SU(2)_R$ 
& $U(1)_Y$
& $SU(3)_c$
& $U(1)_D$
\tabularnewline \midrule 
$\hat{S}_L^{\D_i}$
& 0
& 1/2
& 0
& 1/3
& 3
& -$\qdi$
\tabularnewline 
$\hat{S}_L^{\U_i}$
& 0
& 1/2
& 0
& 1/3
& 3
& -$\qui$
\tabularnewline 
$\hat{S}_R^{\D_i}$
& 0
& 0
& 1/2
& 1/3
& 3
& -$\qdi$
\tabularnewline 
$\hat{S}_R^{\U_i}$
& 0
& 0
& 1/2
& 1/3
& 3
& -$\qui$
\tabularnewline 
$Q^{\D_i}$
& 1/2
& 0
& 0
& 0
& 0
& $\qdi$
\tabularnewline 
$Q^{\U_i}$
& 1/2
& 0
& 0
& 0
& 0
& $\qui$
\tabularnewline \bottomrule \end{tabular} 
\caption{
Spin and quantum numbers of messenger fields and dark fermions.   
$U(1)_D$ is the gauge symmetry associated to dark photon interactions.
}
\label{tab1}
\end{center} \end{table}

Notice that messenger fields also carry the same $U(1)_D$ charges as the associated dark fermions, hence the $U(1)_D$ charge identifies the flavor states.
We report below only the part of the interaction Lagrangian  ${\cal L}^I_{MS}$ relevant to our discussion, referring the reader to \cite{Gabrielli:2016vbb} for the full expression. The scalar mediator interactions responsible for the radiative generation of the diagonal Yukawa entries in the quark sector then are
\bea
{\cal L}^I_{MS} &=&
\hat{g}_L   \left( \sum_{i=1}^{N}\left[\bar{\psi}^i_L Q_R^{\U_i}\right] \hat{S}^{\U_i}_{L} +
\sum_{i=1}^{N}\left[\bar{\psi}^i_L Q_R^{\D_i}\right] \hat{S}^{D_i}_{L}\right)
\nonumber\\
&+&
\hat{g}_R \left(\sum_{i=1}^{N}\left[\bar{\psi}^i_R Q_L^{\U_i}\right] \hat{S}^{\U_i}_{R} +
\sum_{i=1}^{N} \left[\bar{\psi}^i_R Q_L^{\D_i}\right] \hat{S}^{\D_i}_{R}\right) 
\nonumber\\
&+& \;
\lambda \sum_{i=1}^{N}\left(\tilde{H}_L^{\dag} \hat{S}^{\U_i}_L \hat{S}^{\U_i\dag}_R\tilde{H}_R
+ H^{\dag}_L \hat{S}^{\D_i}_L \hat{S}^{\D_i\dag}_RH_R \right) 
\,+\, h.c.  \,,
\label{LagMS}
\eea 
where the color and $SU(2)_{L,R}$ contractions are left understood. The $SU(2)_{L,R}$ doublets $\psi^i_{L,R}=\left(\begin{array}{c} U^i_{L,R}\\D^i_{L,R} \end{array}\right)$
represent here the SM up ($U$) and down ($D$) quark fields,  
$H_{L,R} =\left(\begin{array}{c} H_{L,R}^{\pm}\\H^{0}_{L,R}  \end{array}\right)$
are the Higgs doublets and $\tilde{H}_{L,R}$ are, as usual, defined as
$\tilde{H}_{L,R}=i\sigma_2 H^{\star}_{L,R}$. 
The two constants $\hat{g}_L$ and $\hat{g}_R$ in Eq.~\eqref{LagMS} are flavor-universal parameters required to have perturbative values $\hat{g}_{L,R}< 1$. In the following we also identify $\hat{g}_L=\hat{g}_R=\g$ as imposed by the LR symmetry.

We show in the next subsection how the interactions in Eq.(\ref{LagMS}) give rise to the diagonal Yukawa couplings for the up and down quark fields.

\subsection{The origin of the flavour hierarchy}
After the $SU(2)_L\times SU(2)_R$ breaking, the free Lagrangian of the mediator fields becomes 
\bea
{\cal L}^0_{S}&=& \partial_{\mu} \hat{S}^{\dag} \partial^{\mu}\hat{S} - \hat{S}^{\dag} M^2_S \hat{S},
\eea
where $\hat{S}\equiv (\hat{S}_L,\hat{S}_R)$ and the square mass term is given by
\begin{equation}
M^2_S = \left (
\begin{array}{cc}
m^2_L & \Delta \\
\Delta & m^2_R 
\end{array}
\right),
\label{M2}
\end{equation}
where we omitted the $U,D$ indices of fields since they are not relevant for the purposes of the following discussion. In the above equation, $v_{L,R}$ indicate the vevs of $\hat{H}_{L,R}$, respectively. The messenger mass mixing term $\Delta=\frac{1}{2}\lambda v_R v_L$ instead parametrizes the left-right scalar mixing and originates after SSB from an interaction term in the Higgs sector $\lambda(\hat{H}^{\dag}_L \hat{S}_L) (\hat{S}_R^{\dag} \hat{H}_R)$, needed to generate the Yukawa couplings -- see \cite{Gabrielli:2016vbb} for more details. 

The assumption of a global $SU(6)$ flavor symmetry forces the terms appearing in Eq.~(\ref{M2}) for each $\hat{S}^{\U_i,\D_i}_{L,R}$ flavor component to have the same value. As a consequence, the $M^2_S$ matrix in Eq.~(\ref{M2}) can be separately diagonalized for each flavor via the unitary matrix
\begin{equation}
U = \left (
\begin{array}{cc}
\cos{\theta} & \sin{\theta} \\
-\sin{\theta} & \cos{\theta} 
\end{array}
\right),
\label{U2}
\end{equation}
 with  
$\tan{2\theta}=\frac{2\Delta}{m_L^2-m_R^2}$. 
The eigenvalues of the corresponding diagonal mass matrix $M^{2 \,\rm diag}_S=U M^2_S \,U^{\dag}$ then are
\bea
m^2_{\pm}=\frac{1}{2}\left(m^2_L+m_R^2 \pm
\left[(m^2_L-m^2_R)^2+4\,\Delta^2\right]^{1/2}
\right)\, ,
\eea
and $m^2_L=m^2_R=\bar{m}^2$ in our symmetric LR scenario. The $U$ matrix elements consequently
simplify to $U(i,i)=1/\sqrt{2}, \;U(1,2)=\!-U(2,1)=1/\sqrt{2}$, yielding square-mass eigenvalues 
\bea
m^2_{\pm}=\bar{m}^2(1\pm \xi)\, ,
\eea
where the mixing parameter $\xi$ is given by 
\bea
\xi=\frac{\lambda v_R v_L}{2\bar{m}^2}\, .
\label{mixing}
\eea

\begin{figure}[h]
  \centering
    \includegraphics[width=.7\textwidth]{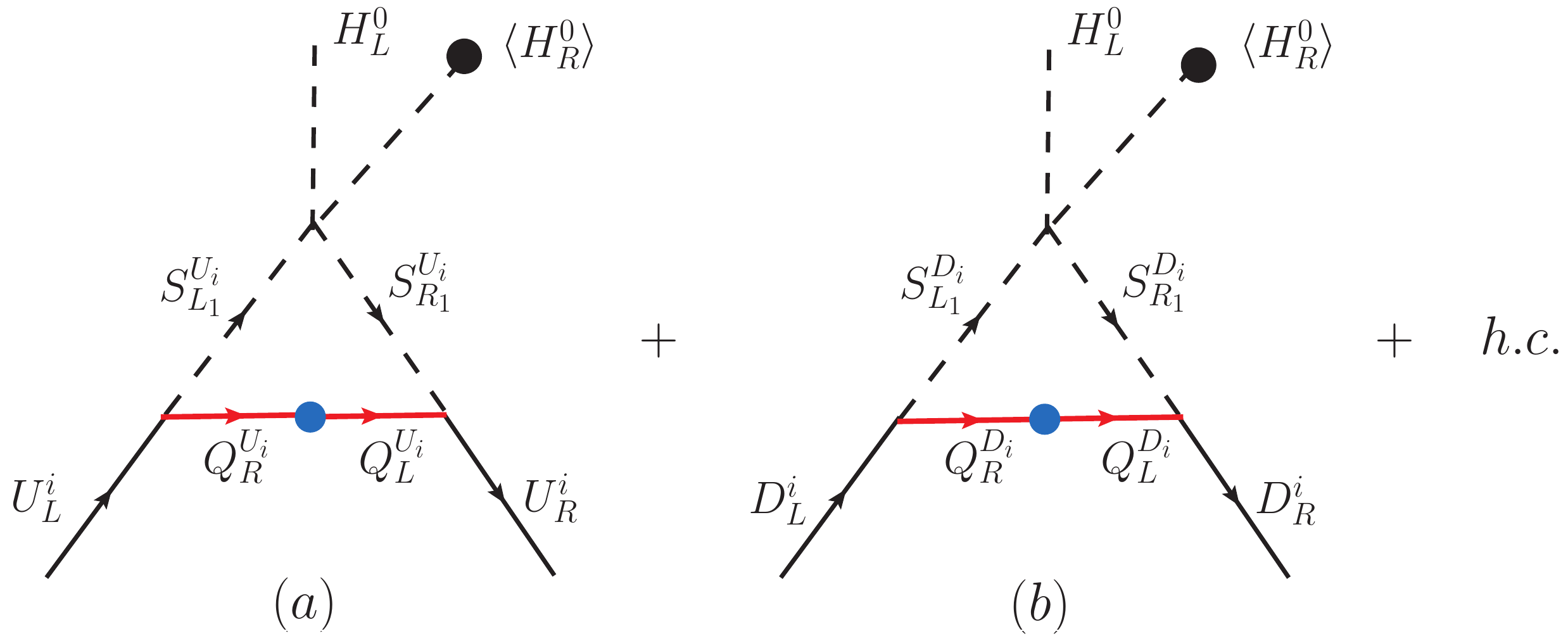}
    \caption{Feynman diagrams responsible for the radiative generation of the  up-type quarks (a) and down-type quarks (b) Yukawa couplings. The dot on an external Higgs line indicates that the field is set to its vev.}
  \label{fig:figd}
\end{figure}

The diagonal entries $Y_f$ of the SM Yukawa matrices, are then generated through the diagrams presented in Fig.\ref{fig:figd}, which at low energies give rise to the dimension 5 operator
\bea
{\cal L}_{eff} &=& \frac{1}{\Lambda^f_{\rm eff}}(\bar{\psi}^f_L H_L)(H^{\dag}_R \psi^f_R) + h.c. \, ,
\label{OYukawa}
\eea
after the replacement $\hat{H}_R\to v_R$. We indicated here with $(\psi^f_L) \, H_L$ and $(\psi^f_R)\, H_R$ the relevant (fermion) Higgs doublets belonging to the indicated chiral sector.
The effective constant $\Lambda^f_{\rm eff}$ is determined by matching the matrix elements obtained from the effective low energy operator in Eq.(\ref{OYukawa}) with the corresponding expressions in the fundamental theory, given by the computation of the one-loop diagram in
Fig.\ref{fig:figd}.
The resulting expression for $Y_f$ as a function of the mixing $\xi$ is
\cite{Gabrielli:2016cut}
\bea
Y_f&=&\left(\frac{\g^2 }{16 \pi^2 }\right)
\left(\frac{\xi m_{Q^f}\sqrt{2}}{v_L}\right) f_1(x_f,\xi)\, ,
\label{Yukexact}
\eea
where $m_{Q^f}$ is the mass of the dark fermion $Q^f$, $x_f=m_{Q^f}^2/\bar{m}^2$, and the loop function $f_1(x,\xi)$ is given by 
\cite{Gabrielli:2016cut}
\bea
f_1(x,\xi)&=&\frac{1}{2}\left[
C_0(\frac{x}{1-\xi})\frac{1}{1-\xi}+C_0(\frac{x}{1+\xi})\frac{1}{1+\xi}\right]\,,
\label{f1}
\eea with
\bea
C_0(x)=\frac{1-x\left(1-\log{x}\right)}{(1-x)^2}\, .
\label{C0}
\eea 
The dark fermion masses $m_{Q^f}$ can be generated non-perturbatively in the dark sector via the Lee-Wick mechanism for chiral symmetry breaking \cite{Gabrielli:2007cp}. Due to the presence of the higher-derivative Lee-Wick term in the $U(1)_D$ gauge sector \cite{Lee:1971ix,Lee:1970iw,Grinstein:2007mp}, the dark fermions acquire masses given by \cite{Gabrielli:2013jka}
\bea
m_{Q^f}=\Lambda \exp{\left(-\frac{\gamma}{q^2_f\alpha_D}\right)}\, ,
\label{DFmass}
\eea
where $q_f$ are the different ${\cal O}(1)$ $U(1)_D$ charges of dark fermions and $\gamma$ is a constant connected to an anomalous dimension. The scale $\Lambda$ refers to the scale in the Lee-Wick term of the $U(1)_D$ gauge sector. By inserting Eq.(\ref{DFmass}) into (\ref{Yukexact}), we can then see that the hierarchy in the diagonal components of the obtained Yukawa coupling matrix simply reflects the exponential spread of the dark fermion masses.

\section{The origin of the CKM matrix}

In the previous Section we discussed the generation of hierarchical diagonal terms in the Yukawa matrix. In order to fully address the problem of flavour, we need now to address the remaining off-diagonal entries, which are responsible for inducing the quark mixing modelled in the CKM matrix. 

These terms were modelled in the original framework by generalizing the interactions in Eq.(\ref{LagMS}) to a mixing between quarks of generation $i$ and dark fermions and messengers of generation $j$ \cite{Gabrielli:2016vbb}, postulated to be small in accordance with the MFV hypothesis. In order to address this shortcoming, in the following we detail a new dynamical mechanism to generate suppressed off-diagonal terms in the Yukawa matrix. We then introduce a complex single scalar field $F$, the dark flavon, which transforms under $U(1)_D$ with a charge $q_F$. Moreover, we assume that the $U(1)_D$ charges of the dark fermions satisfy the following relations 
\bea
q_F&=&q^{\U}_2-q^{\U}_1=q^{\U}_3-q^{\U}_2=q^{\D}_2-q^{\D}_1=q^{\D}_3-q^{\D}_2\, ,
\label{Qrel}
\eea 
where $q^{\U,\D}_i$ correspond to the $U(1)_D$ charges of the dark fermion $Q^i$ associated to the quark generation $i$ in the up $(U)$ and down $(D)$ sectors, respectively. It then follows that $q^{\U,\D}_2=(q^{\U,\D}_3-q^{\U,\D}_1)/2$, and gauge invariance imposes analogous relations for the $U(1)_D$ charges of messenger fields $S^{\U,\D}_{L_i}$\footnote{Notice that the relations in Eq.(\ref{Qrel}) could be satisfied by simply considering charges taken from a sequence of integer numbers.}. 

With this setup, the Lagrangian of the model can be extended to the following interactions that source the off diagonal terms in the Yukawa matrices of up and down-type quarks:
\bea
\label{eq:LagF}
{\cal L}_{F}&=&{\cal L}_0(F) + \left(\eta_{\U} \bar{Q}^{\U_1} Q^{\U_2}F+\eta_{\U}^{\prime}\bar{Q}^{\U_2} Q^{\U_3}\right)F + \left(\eta_{\D} \bar{Q}^{\D_1} Q^{\D_2}+\eta_{\D}^{\prime}\bar{Q}^{\D_2} Q^{\D_3}\right)F
\\
\nonumber
&+&\mu_U\left(\hat{S}^{U_1\dag}_L \hat{S}^{U_2}_L+\hat{S}^{U_2\dag}_L \hat{S}^{U_3}_L\right) F +\mu_D\left(\hat{S}^{D_1\dag}_L \hat{S}^{D_2}_L+\hat{S}^{D_2\dag}_L \hat{S}^{D_3}_L\right) F +\{L\leftrightarrow R\}
+ h.c.
\eea
where ${\cal L}_0(F)$ is the free Lagrangian of the dark flavon field $F$.

\begin{figure}[h]
  \centering
    \includegraphics[width=.9\textwidth]{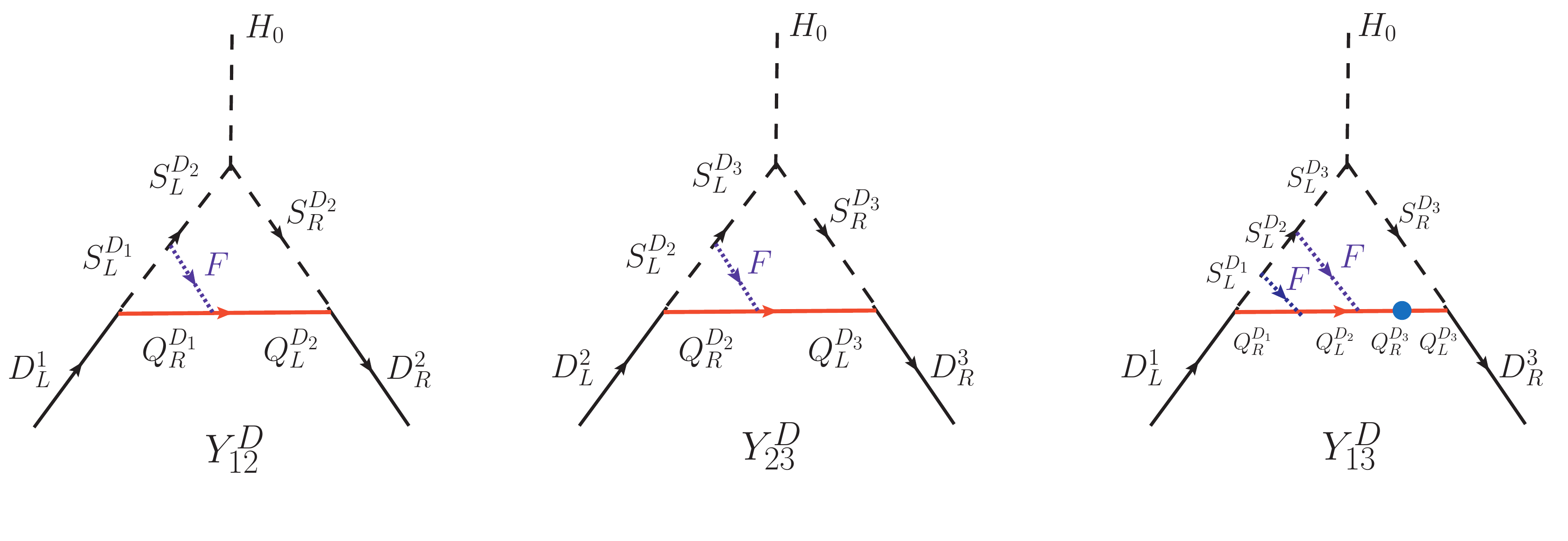}
  \caption{Diagrams responsible for the radiative generation of the off-diagonal terms in the Yukawa couplings for the down-type quarks. An external $<H_R>$ vacuum expectation value insertion at the Higgs vertex is understood. Similar diagrams source the off-diagonal terms in the up-type quark Yukawa matrix.}
  \label{fig2}
\end{figure}

By using the interactions in Eq.(\ref{eq:LagF}), we can generate the desired terms via the single or double exchange of the $F$ field in the same one-loop diagram responsible for the diagonal terms, as shown in Fig.\ref{fig2}. The resulting off-diagonal components are then naturally smaller than the diagonal elements, and consequently induce only negligible corrections to the mass hierarchy regulated by the latter. We anticipate that at least two different couplings, $\eta_{\U,\D}$ and $\eta_{\U,\D}^{\prime}$ of the $F$ field to the dark fermions (or, alternatively, to the messenger fields), are needed in order to source the CP-violating phase in the resulting CKM matrix. 

Following the structure of the diagrams in Fig.\ref{fig2}, we can see that the Yukawa matrices generated are necessarily hermitian and possess a well defined structure, given in terms of the effective parameters by
\bea
\frac{Y^{\U}}{y_t} \sim 
\begin{pmatrix}
  A_r^{\U} \rho_u & \epsilon^{\U}_{1}  & A^{\U}\epsilon^{\U}_{1}\epsilon^{\U}_{2}\\ 
\epsilon^{\U*}_{1} & \rho_c & \epsilon^{\U}_{2}\\ 
A^{\U}\epsilon^{\U*}_{1}\epsilon^{\U*}_{2}  & \epsilon^{\U*}_{2} & 1
\end{pmatrix}\, ,
\eea
and analogously for the down-type Yukawa
\bea
\frac{Y^{\D}}{y_b} \sim 
\begin{pmatrix}
A_r^{\D} \eta_d & \epsilon^{\D}_{1}  & A^{\D}\epsilon^{\D}_{1}\epsilon^{\D}_{2}\\ 
\epsilon^{\D*}_{1} & \eta_s & \epsilon^{\D}_{2}\\ 
A^{\D}\epsilon^{\D*}_{1}\epsilon^{\D*}_{2}  & \epsilon^{\D*}_{2} & 1
\end{pmatrix}\, ,
\eea
where $\rho_i=m_i/m_t$, $\eta_i=m_i/m_b$, whereas $y_b$ and $y_t$ are the bottom and top quark SM Yukawa couplings. The terms
$A^{\D,\U}$ and $A_r^{\D, \U}$ are real coefficients of order ${\cal O}(1-10)$ and ${\cal O}(0-1)$, respectively, while the coefficients $\epsilon^{\U,\D}_{1,2} \ll 1$. The coefficient $A^{\D, \U}$, in particular, is meant to compensate extra loop and coupling suppression obtained in modelling a three-loop contribution as a product of two two-loop contributions. Differently, the factor $A_r^{\D, \U}$ weights the contribution of quark mixing into the first generation quark masses. With the expressions above, we can then compute the CKM matrix as 
\begin{equation}
  V_{CKM} = V^{\U} (V^{\D})^\dg
\end{equation} 
where $V^{\U}$ and $V^{\D}$ are the unitary matrices that diagonalize the corresponding Yukawa matrices
\begin{equation}
  Y^{x}_{diag} =  V^x Y^x {(V^x)}^\dg
\end{equation}
with $x={\U}, {\D}$.

The value of the parameters and the phases in the two unitary matrices can then be determined by matching the obtained CKM matrix elements with the corresponding experimental values. 

\subsection{Matching the observed quark mixing and hierarchy}

In order to make contact with measurements we perform a numerical scan of the resulting CKM matrix, selecting the parameters that reproduce the observed quark mixing and mass hierarchy. 

In this first analysis we simplify our task by neglecting all phases, which can be straightforwardly matched by arranging the relative phases of the complex parameters that enter the diagrams discussed in the previous Section. Furthermore, we assume that the Lagrangian of the model is already given on a field basis where the up-type quark Yukawa couplings are diagonal. This choice introduces mixing factors in the interactions of up-type quarks with mediators and dark fermions, but noticeably simplifies our investigation that now needs to keep track of the effective parameters in $Y^{\D}$ only.

\begin{table}[t]
  \centering
  \begin{tabular}{c| c c c c}
  \toprule 
   Parameter &  Lower bound &  Upper bound& Sampling& Number of points  \\
   \midrule
   $A_r^{\D}$ & $10^{-5}$  & 10 & exp$_{10}$ & 200 \\
   $\epsilon_1^{\D}$ & $10^{-2.5}$  & $10^{-2}$ & exp$_{10}$ & 200 \\
   $\epsilon_2^{\D}$ & $10^{-1.5}$  & $10^{-1}$ & exp$_{10}$ & 200 \\
   $A^{\D}$ & $5$  & 45 & lin & 200 \\
   \bottomrule
  \end{tabular}
  \caption{The effective parameter space selected after a preliminary scan.}
  \label{table:par}
\end{table}

We report in Table \ref{table:par} the ranges of the relevant parameters considered in the analysis, selected after a preliminary scan. For every combination of the parameters drawn from these intervals, we diagonalize the resulting Yukawa coupling matrix and obtain the masses of the down-quark masses upon multiplication by a factor of $m_b$, as well as the mixing angle of the CKM matrix. We then compare our findings to the corresponding experimental values, retaining only the combination of input parameters that result in observables within the acceptance limits reported in Table~\ref{tab:accept}.
For the light quark masses we adopt conservative ranges to account for the uncertainty caused by the running of these parameters. For the remaining quantities, instead, we use the $3\sigma$ limit from Ref.~\cite{Eidelman:2004wy}.

\begin{table}[h]
  \centering
  \begin{tabular}{c| c c}
  \toprule 
   Parameter &  Lower bound &  Upper bound  \\
   \midrule
   $m_d$ (GeV) & $10^{-3}$  & $10^{-2}$ \\
   $m_s$ (GeV) & $5 \times 10^{-2}$  & $1.5 \times 10^{-1}$ \\
   $m_b$ (GeV) & $4.06$  & $4.3$ \\
   $\sin\theta_{12}$ & 0.2195 & 0.2291 \\
   $\sin\theta_{23}$ & 0.0368 & 0.0458 \\
   $\sin\theta_{13}$ & 0.0022 & 0.0052\\
   \bottomrule
  \end{tabular}
  \caption{The acceptance criterions used in the scan.}
  \label{tab:accept}
\end{table}

\begin{figure}[h]
  \centering
    \includegraphics[width=.4\textwidth]{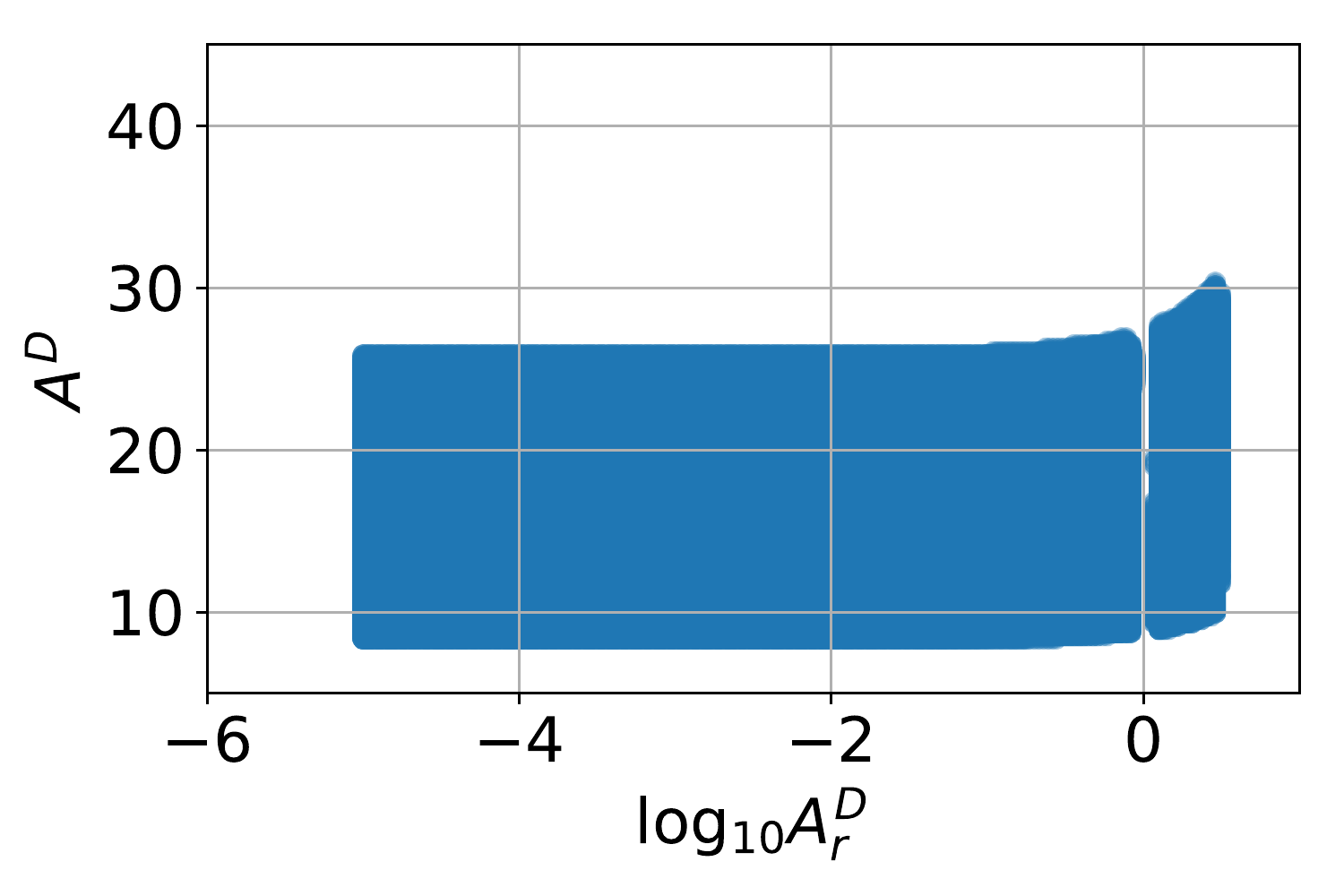}
    \includegraphics[width=.4\textwidth]{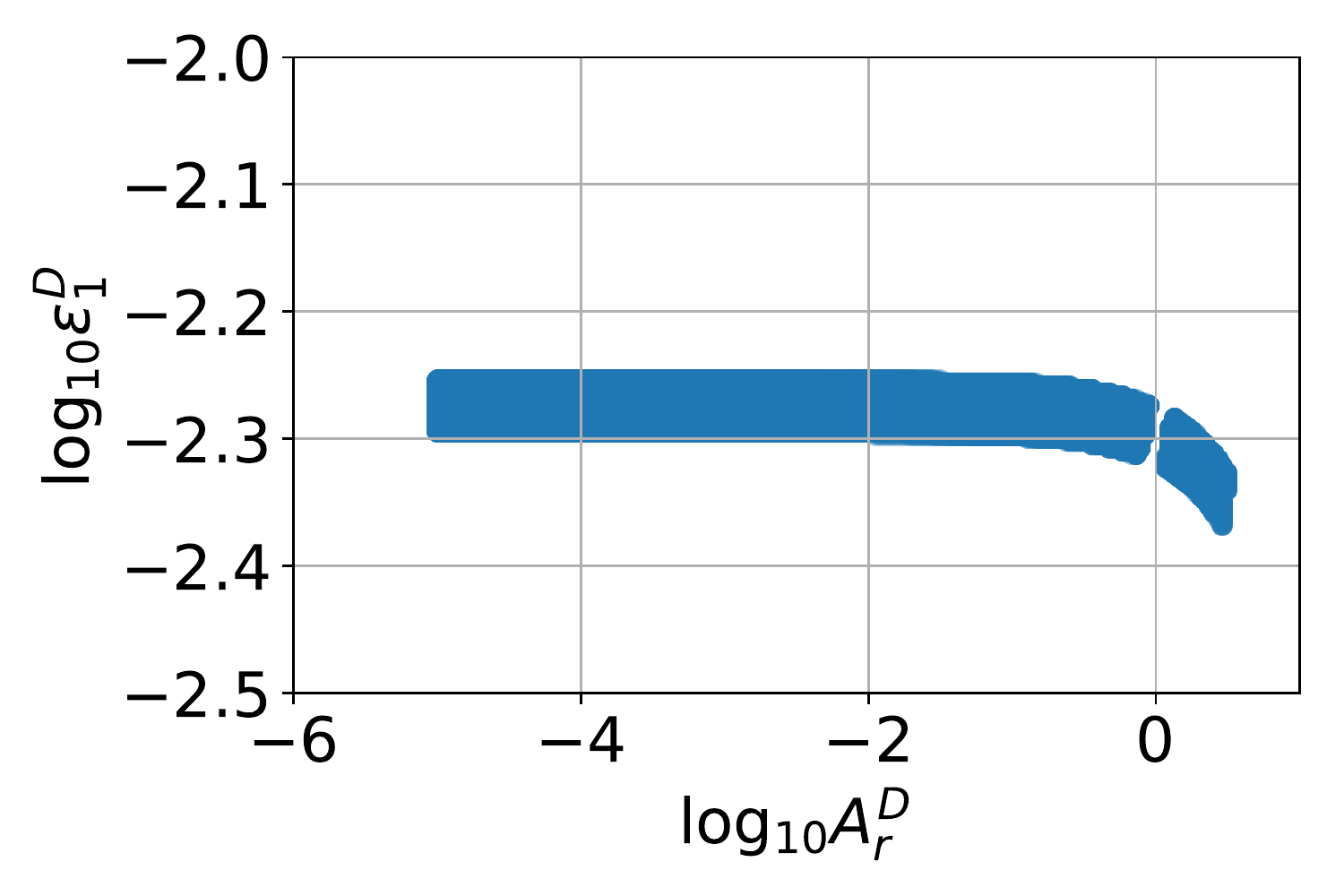}
    \\
    \includegraphics[width=.4\textwidth]{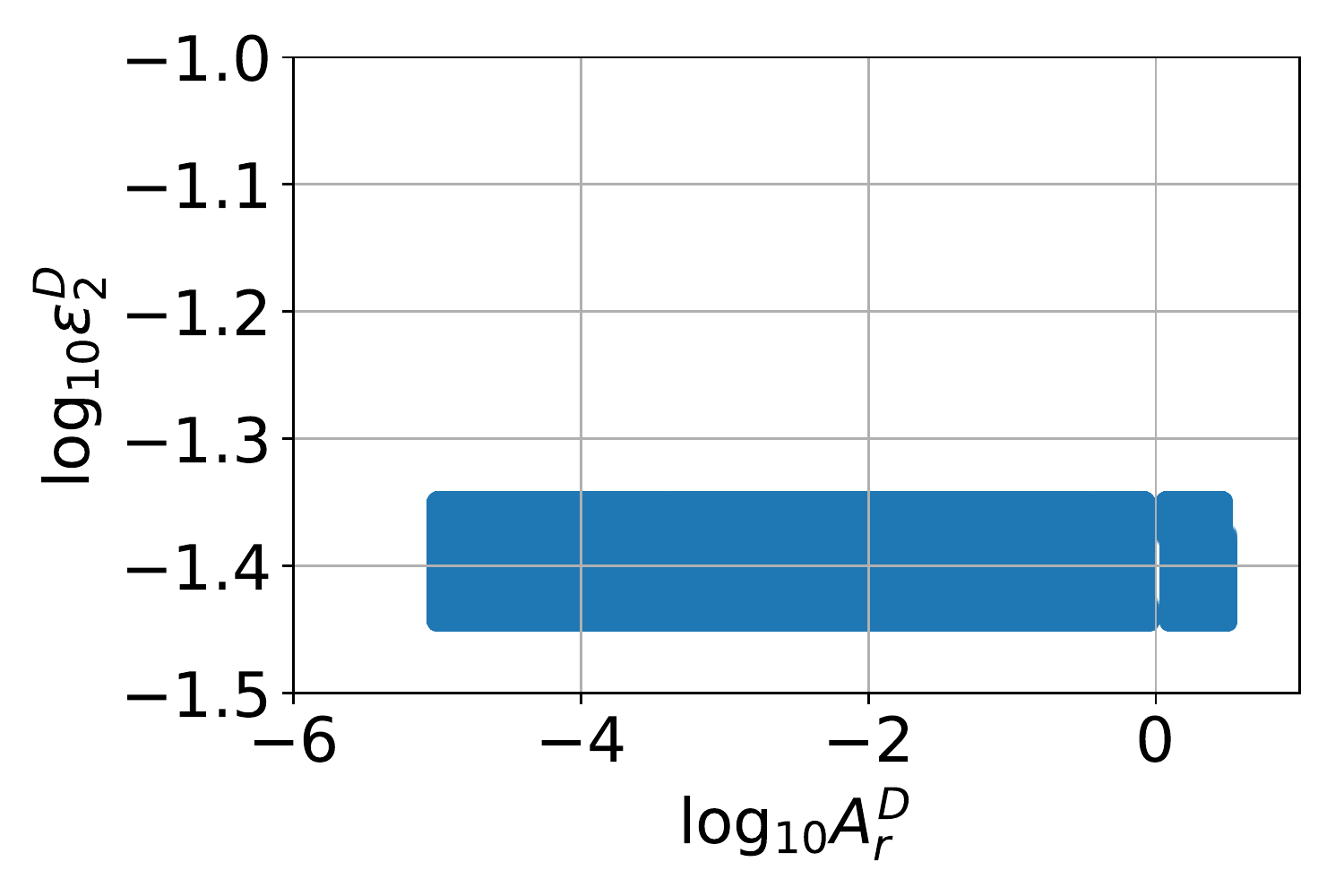}
    \includegraphics[width=.4\textwidth]{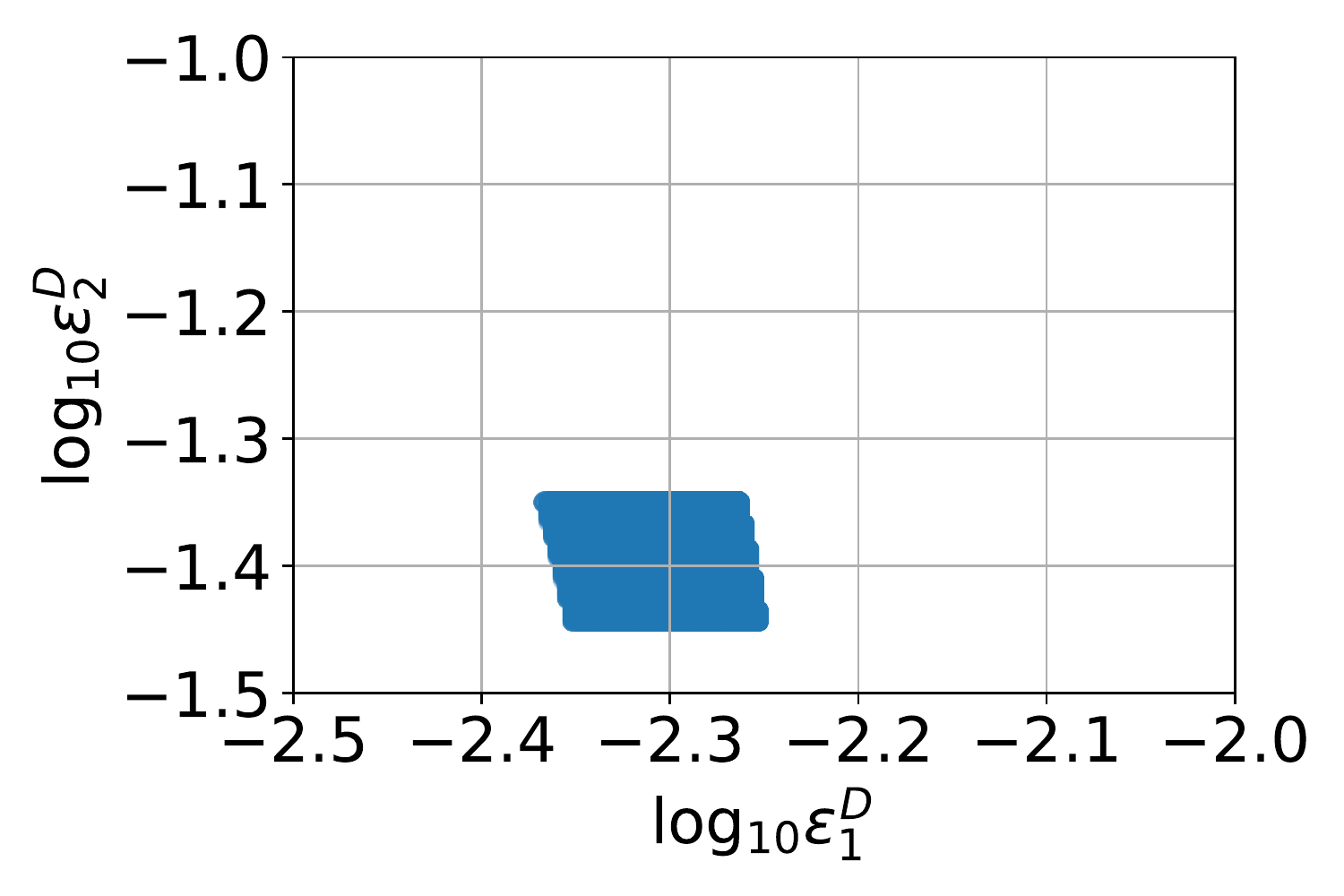}
    \\
    \includegraphics[width=.4\textwidth]{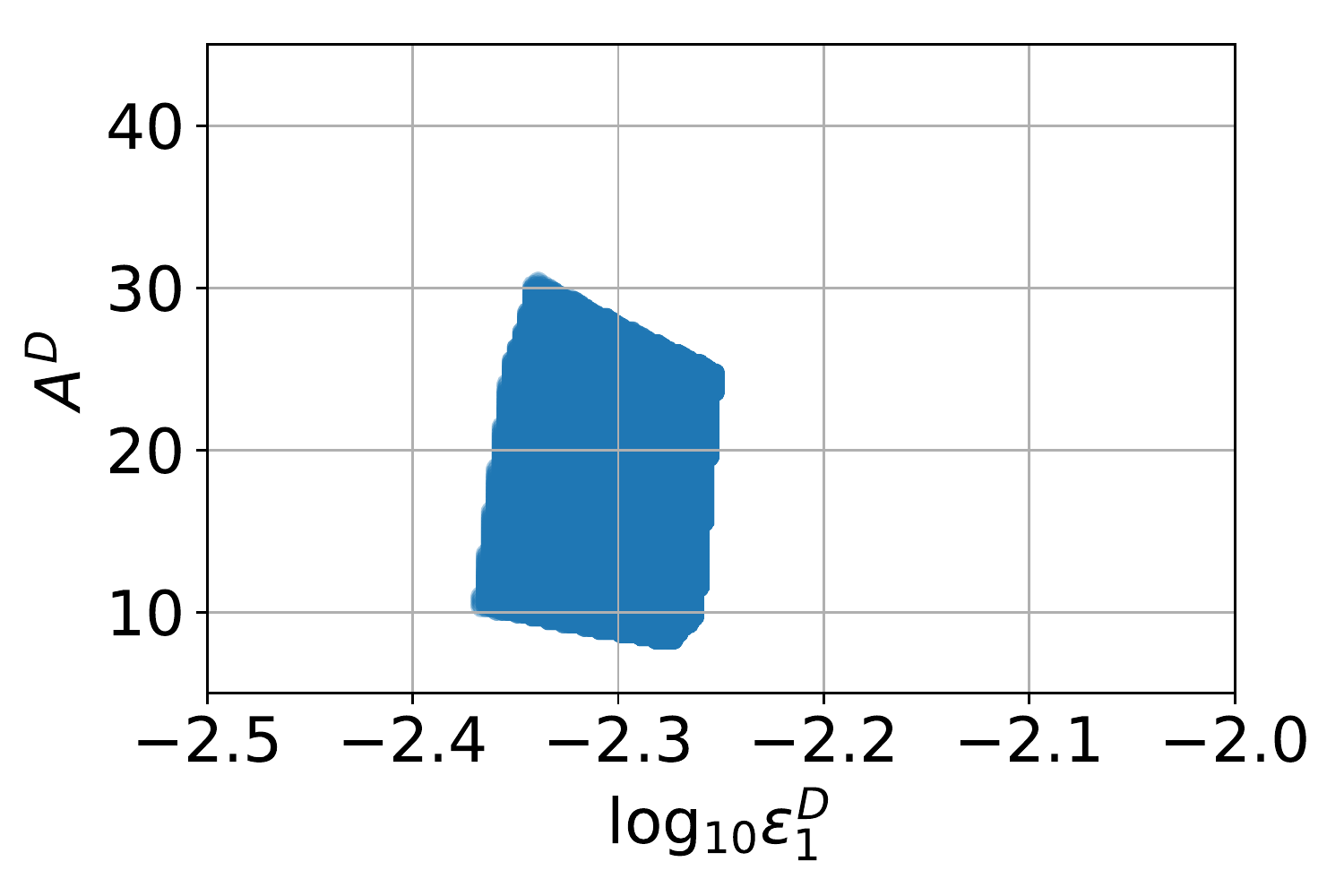}
    \includegraphics[width=.4\textwidth]{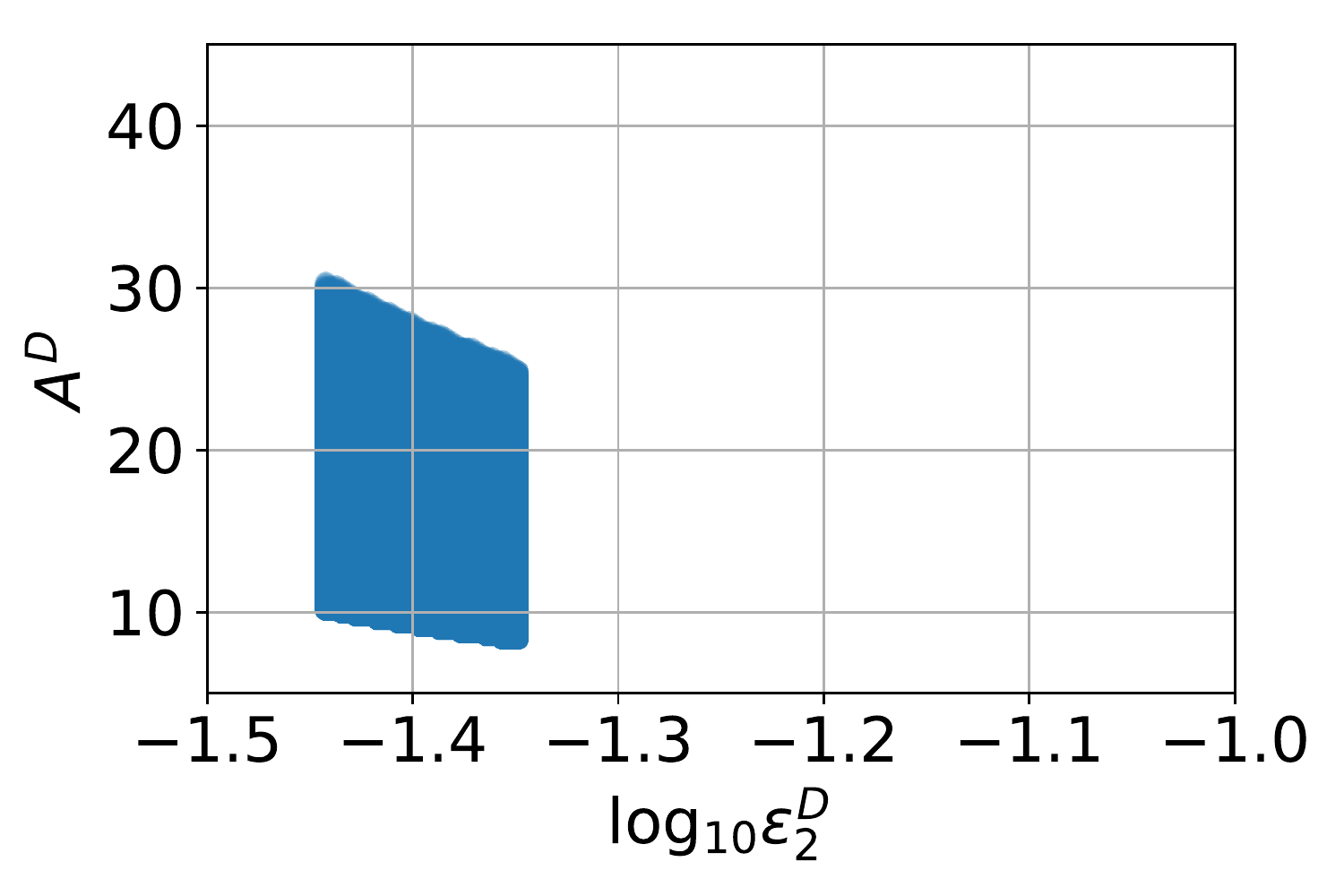}
  \caption{The combinations of effective input parameters selected by the acceptance intervals given in Table~\ref{tab:accept}.}
  \label{Fig:input_par}
\end{figure}

The region of the effective parameter space selected by our methodology is presented in Figure~\ref{Fig:input_par}. The first three panels highlight the dependence of the selected effective parameters on $A_r^{\D}$ and reveal the presence of two distinguished solutions, corresponding to $A_r^{\D}<1$ and $A_r^{\D}>1$. In the former case the CKM mixing dominates the contributions into $m_d$, whereas for $A_r^{\D}>1$ this parameter is sourced by the (1,1) component of the Yukawa matrix. The remaining panels show instead the  magnitude of the two two-loop level contributions in Fig.\ref{fig2} required to match the observations, as well as the size of the compensation factor in the employed ansatz for the remaining three-loop contribution.

\begin{figure}[h]
  \centering
    \includegraphics[width=.4\textwidth]{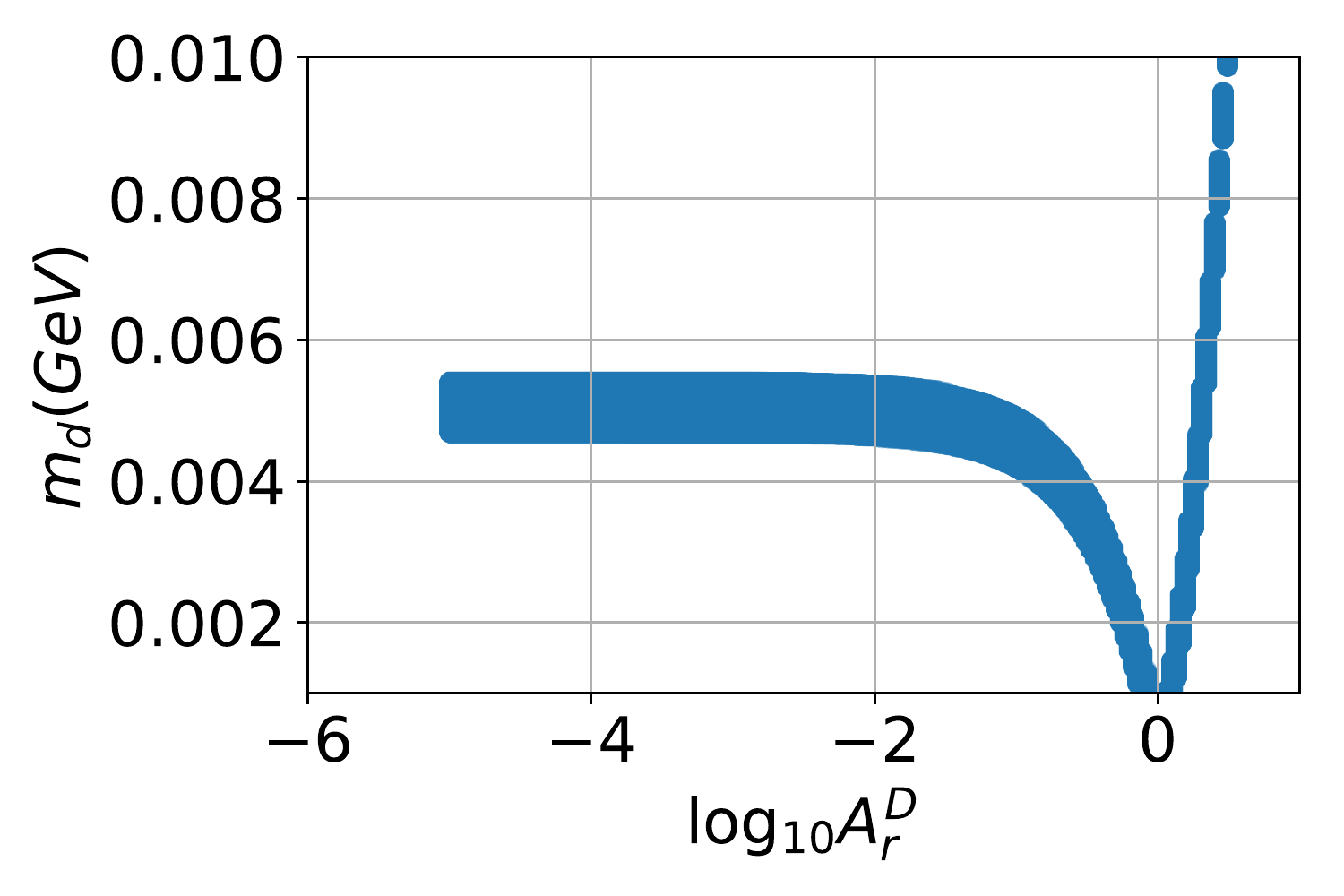}
    \includegraphics[width=.4\textwidth]{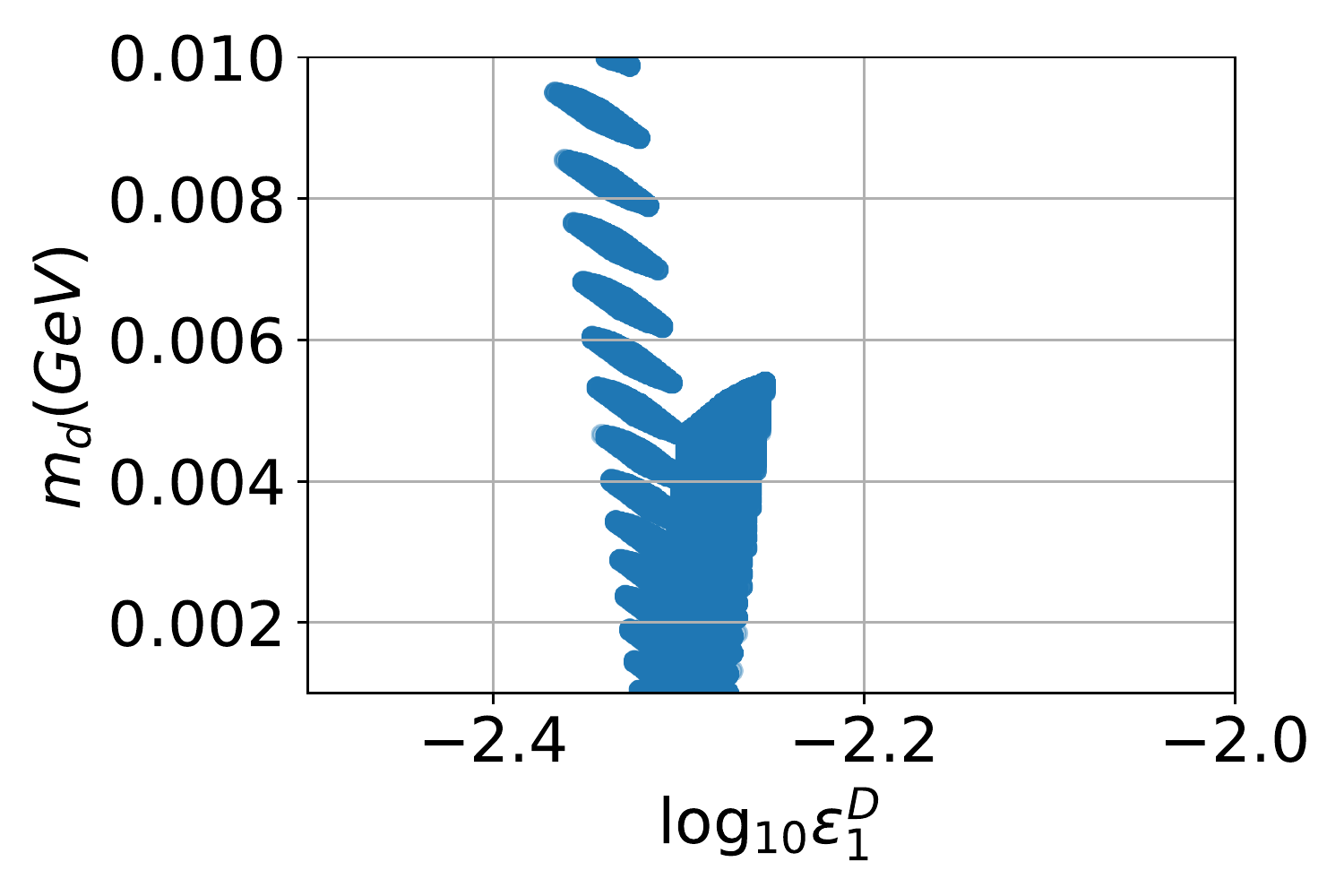}
    \\
    \includegraphics[width=.4\textwidth]{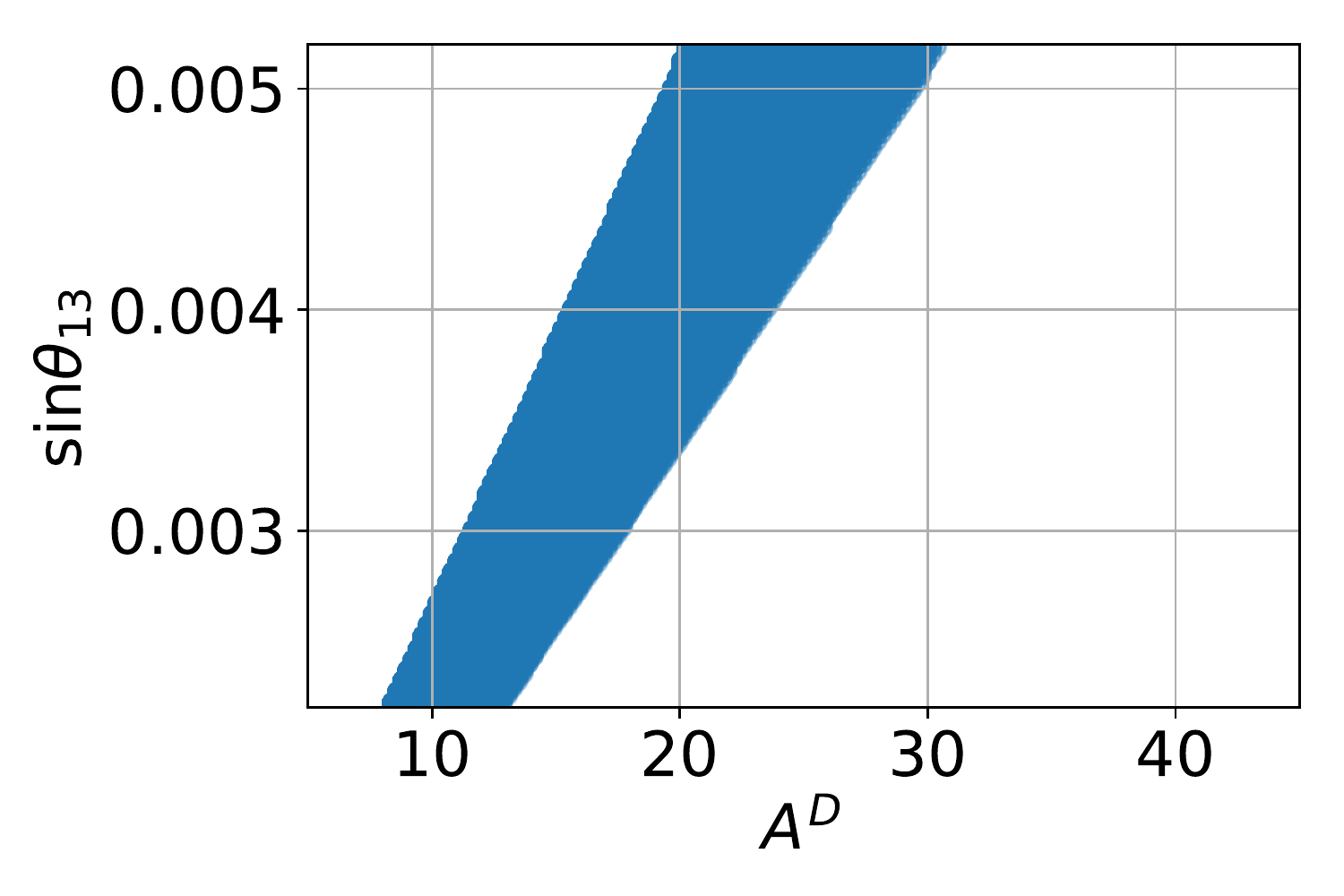}
    \includegraphics[width=.4\textwidth]{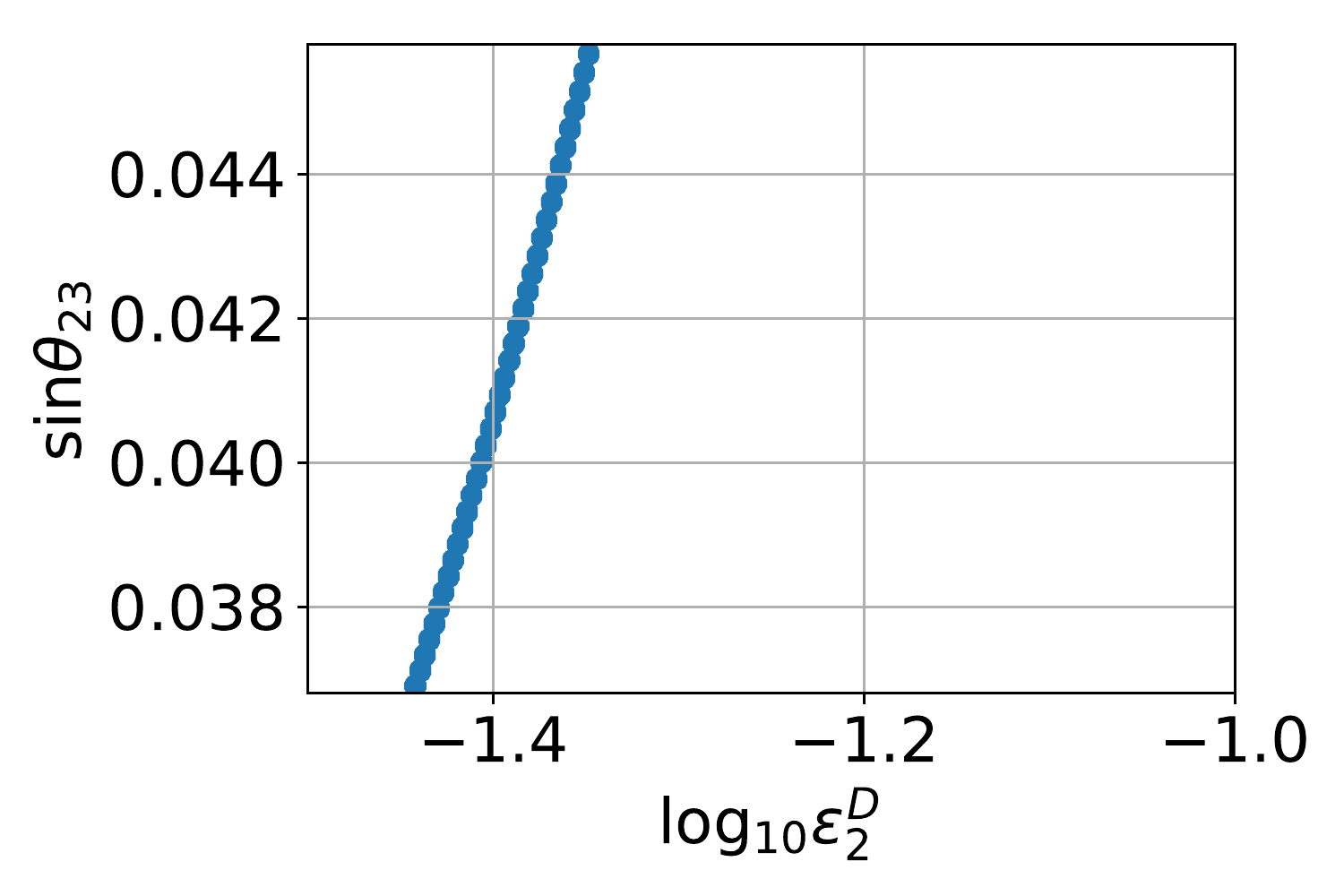}
    \caption{Selected correlation amongst the down-quark masses, the CKM parameters and the effective parameters.}
  \label{Fig:corr}
\end{figure}

As for the correlations between effective input parameters and quark sector observables highlighted by our analysis, we presented in Fig.~\ref{Fig:corr} the most significative trends. The first panel in the top row shows the dependence of the down quark mass on the $A_r^{\D}$ parameter, with the two regimes discussed before being clearly visible. In particular, it is evident that the values of $m_d$ away from 5 MeV require a non-vanishing (1,1) element in the down quark Yukawa coupling matrix. Likewise, as shown in the second panel, large values of this parameter force the loop contribution that regulates the (1,2) entry to the lowest acceptable values. In the bottom row we show instead the correlation between the mixing angles involving the third generation and the corresponding entries in the CKM matrix. Correctly, the former grows with the magnitude of the latter.      

\subsection{Matching with the two and three-loop effective operators from the fundamental theory}

The proposed radiative mechanism naturally presents the propagation of heavy virtual states that mark a clear hierarchy of characteristic mass scales in the model. For this reason, the involved multi-loop amplitudes of Fig.~\ref{fig2} can be given a simple analytical form in terms of an expansion in the mass scale ratios entering the considered loop diagrams.
To avoid the generation of spurious infrared divergences caused by a naive use of a Taylor expansion inside the virtual momenta integration, we employ a series of theoretical tools developed to such purpose, such as the region expansion~\cite{Beneke:1997zp} or the asymptotic large mass expansion~\cite{Smirnov:1994tg}. The latter, in particular, is more suitable for a diagrammatic interpretation in terms of a sum of simpler sub-diagrams that, in many cases, are single scale tadpoles. 
To tackle this theoretical challenge, we employ the {\tt q2e} and {\tt EXP}~\cite{Seidensticker:1999bb,Harlander:1997zb} packages, invoked via {\tt QGRAF}~\cite{Nogueira:1991ex} in order to correctly pick the relevant sub-diagrams and deal with the momentum distribution accordingly. Finally, all the relevant terms are easily computed via the {\tt FORM}~\cite{Ruijl:2017dtg} package {\tt MATAD3}~\cite{Steinhauser:2000ry}. 

The use of the asymptotic expansion in our matching procedure is easily justified by the zero-momentum limit of the external states, as well as by the choice of three different mass scales associated to the propagating dark flavon (with mass $m_F$), the dark messenger (characterized, in first approximation, by a common mass scale $m_S$) and the lightest propagating dark fermion of mass $m_Q$. Whereas the requirement that dark fermion constitute a viable dark matter candidate forces $m_S> m_Q$, the phenomenology of the scenario does not  indicate the ordering of the rest of the spectrum. In particular, although the involved loop functions are certainly sensitive to the relative hierarchy between dark flavons and messengers, a priori we have no reasons to discriminate between the cases $m_F \lessgtr m_S$. 

In order to show how the diagrams in Fig.~\ref{fig2} depend on the involved scales and loop structure, we specialize our discussion to the case $m_F = m_S$ and use a unique parameter $\xi$ to address all the dimensionless couplings. In turn, for the flavon trilinear interaction we use the parametrization $\mu_D =  \lambda\, m_F $.  

In this way, the result of the asymptotic expansion can be expressed in terms of an overall heavy scale $m_F = m_S$, the ratio $x_Q = m_Q/m_F$, and of the loop suppression factor $L = 16\pi^2$. The number of terms involved grows quickly already after the first leading order contribution. For the sake of completeness, we give here the expansion up to eighth order in $x_Q$ as provided by {\tt MATAD3}
\bea
\label{eq:2loopi}
&& \frac{m_F}{v_R}\frac{L^2}{\xi^4 \lambda} Y^{\D}_{12} \sim \frac{m_F}{v_R}\frac{L^2}{\xi^4 \lambda} Y^D_{23} = \nonumber \\ 
&& \frac{1}{12} \left(-27 S_2+\pi ^2-6\right) + \frac{1}{4} \left(45
   S_2-\pi^2+2\right) x_Q^2 + x_Q^4 \left(45 S_2+\frac{11 \log(x_Q)}{3}-\frac{\pi^2}{2}-\frac{179}{36}\right)  + \nonumber \\
&& x_Q^6 \left(99 S_2+4 \log^2(x_Q)+\frac{353 \log(x_Q)}{30}-\frac{\pi ^2}{3}-\frac{17353}{900}\right) + \nonumber \\
&& x_Q^8 \left(\frac{693S_2}{4}+18 \log^2(x_Q)+\frac{976 \log (x_Q)}{35}+\frac{3 \pi^2}{4}-\frac{911367}{19600}\right) + O(x_Q^{10})
\eea
and an analogous expression holds for $Y^{\D}_{23}$. For the three-loop function we have instead
\bea
&&\frac{m_F}{v_R}\frac{L^3}{\xi^6 \lambda} Y^{\D}_{13} = -\frac{1}{24} x_Q \left(459 S_2-112 \zeta(3)+\pi ^2\right) + \nonumber \\
&& - x_Q^3 \frac{
   \left(-9120 \, O^{S_2}_{\epsilon}+15 (196992\, S_2\,+ \,912 \, T^1_{\epsilon}\,-\,45432 \zeta(3)\,+\,7655)\right)}{8640} + \nonumber \\
&& - x_Q^3 \frac{
   \left(4320\,
   \left(-27 \, S_2 + \pi^2 - 6\right) \log(x_Q) + 361 \pi^4 + 30165 \pi^2 \right)}{8640} + \nonumber \\
&& - x_Q^5 \frac{ \left(-98400\, O^{S_2}_{\epsilon} + 1080 \left(846 S_2 + 12\,\pi^2 - 203 \right) \log(x_Q) \right)}{38880} + \nonumber \\
&& - x_Q^5 \frac{ \left(27102114 \, S_2 + 147600\, T^1_{\epsilon}-4948920\, \zeta(3) + 3895\, \pi^4 + 228981\, \pi^2 + 782769 \right)}{38880} + \nonumber \\   
&& x_Q^7 \frac{ \left(-175778400\, O^{S_2}_{\epsilon} + 60480\,
   \log(x_Q) \left(-224325\, S_2 - 10260 \log (x_Q) + 2565\, \pi^2 + 11234\right)\right)}{24494400} + \nonumber \\
&& x_Q^7 \frac{ \left(-1252502190\,
   S_2 + 263667600\, T^1_{\epsilon} - 3233475000\, \zeta(3) + 6957895 \pi^4 \right)}{24494400} + \nonumber \\
&& x_Q^7 \frac{ \left(740318925\, \pi^2 + 3458071731\right)}{24494400} + O(x_Q^9)
\eea
In both the above expressions $v_R$ indicates the vev of $H_R$ and notations for loop functions are taken from \cite{Steinhauser:2000ry}.

\begin{figure}[h]
\centering
\includegraphics[width=.45\linewidth]{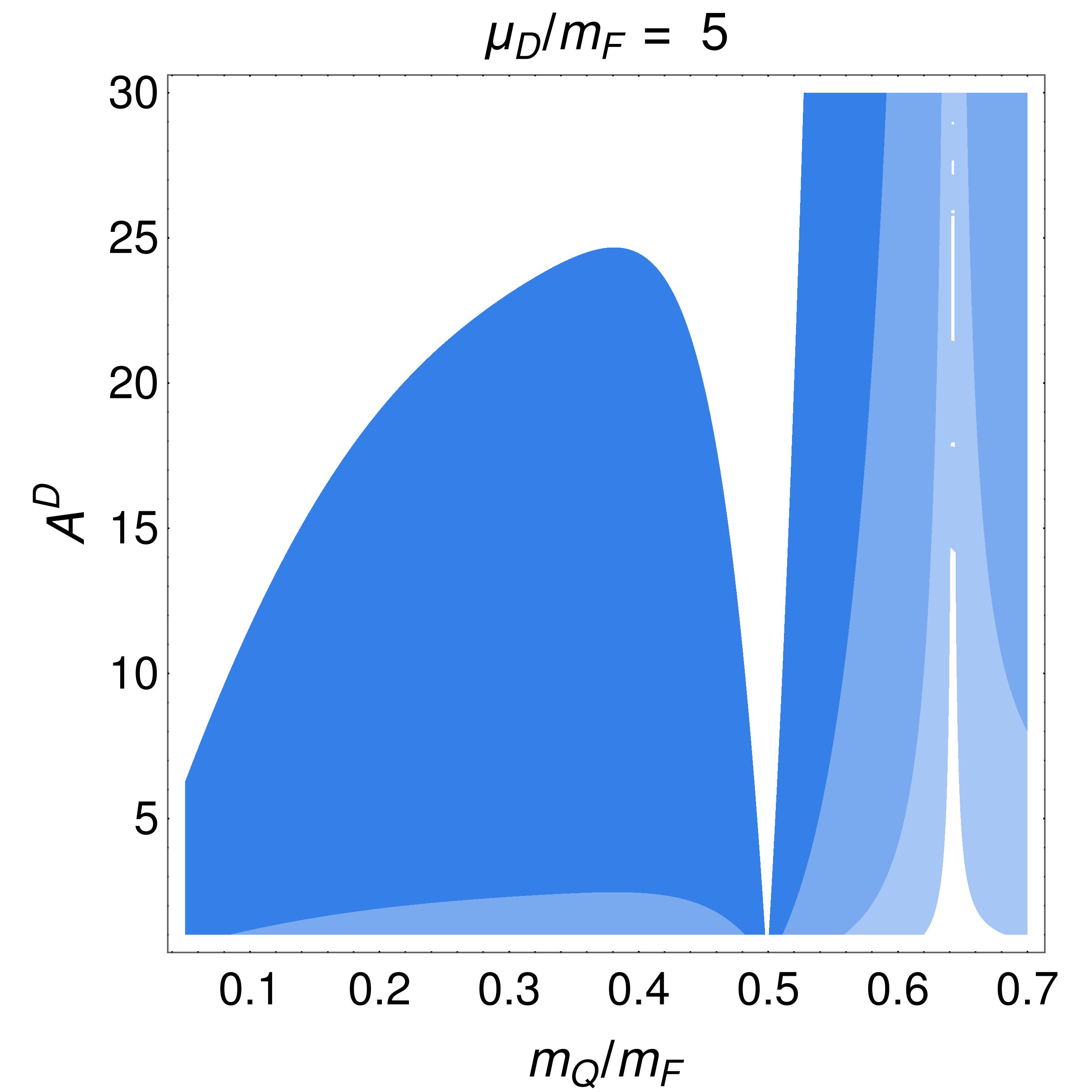}
\includegraphics[width=.45\linewidth]{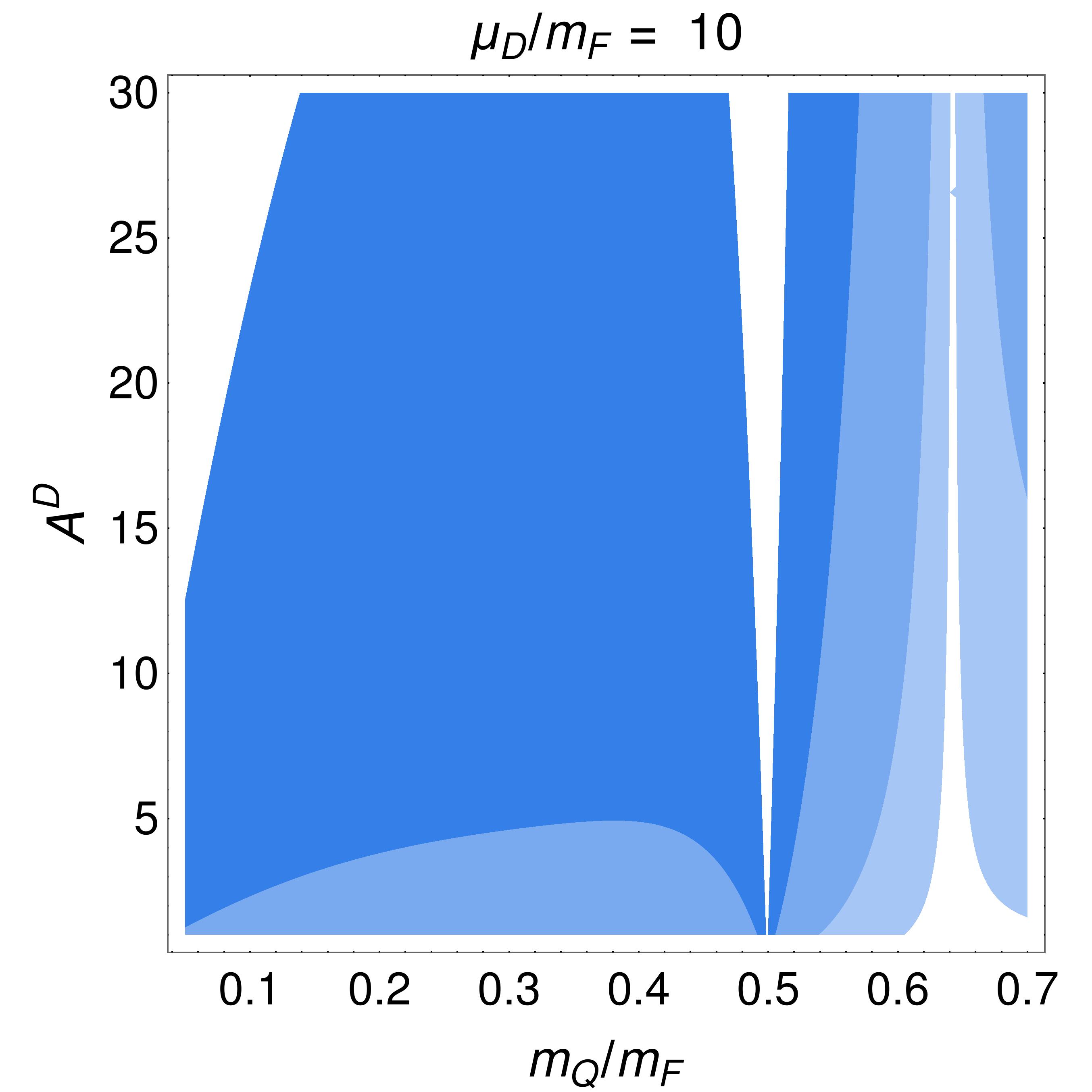}\\
\includegraphics[align=c, width=.45\linewidth]{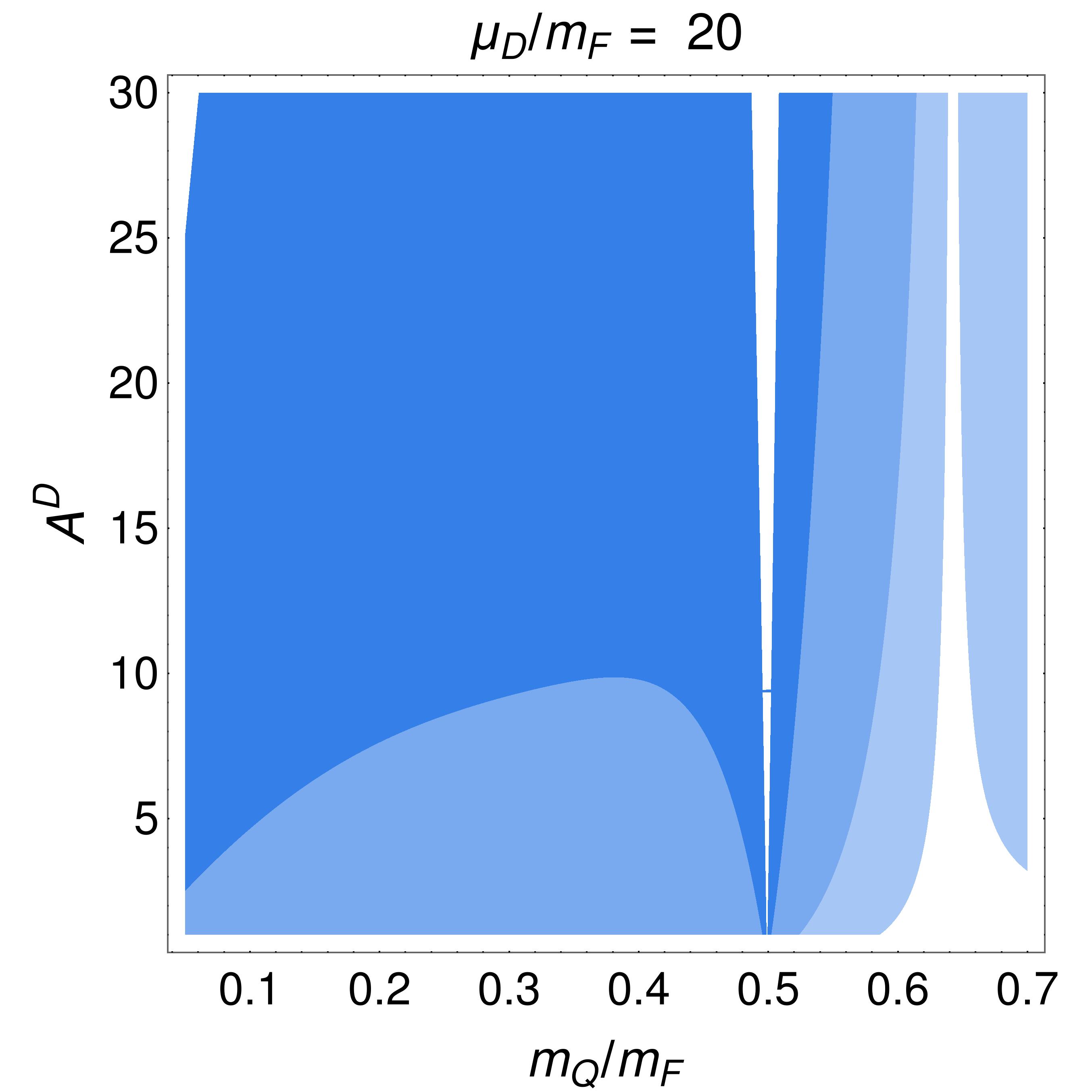}
\includegraphics[align=c, width=.45\linewidth]{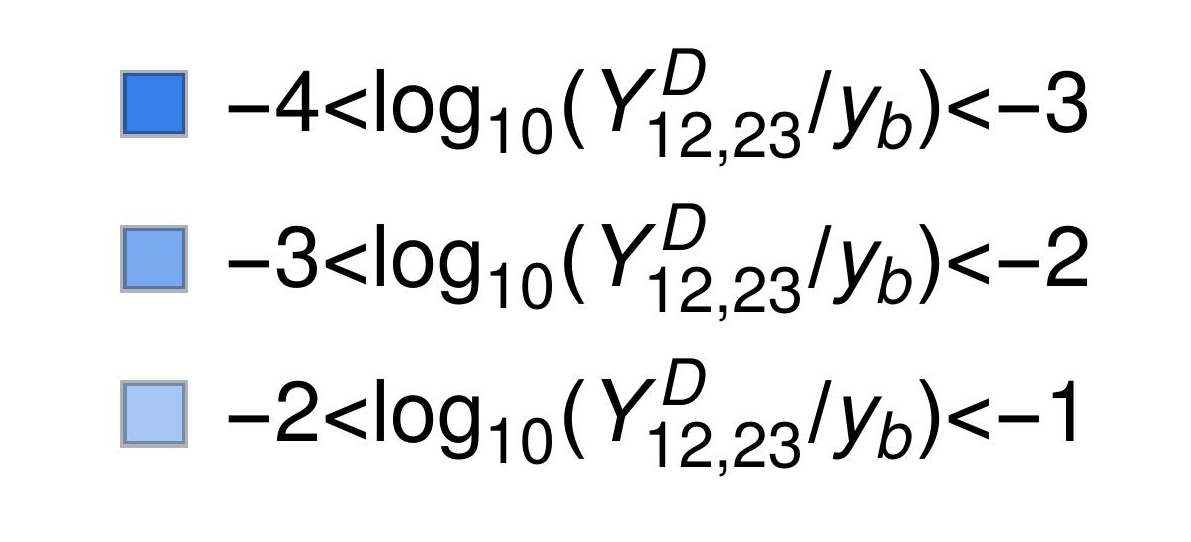}
\caption{Values of the two-loop function (rescaled by the bottom Yukawa coupling) obtained with the analytical expression in Eq.~\eqref{eq:2loopi}, as a function of the $A^D$ parameter and of the mass hierarchy between messengers and dark fermions. All dimensionless coupling constants are set to $1/2$, while the dark flavon trilinear coupling is set at the reported values.}
\label{fig:loopi}
\end{figure}

\FloatBarrier

We use the above expression to evaluate the two-loop contribution, shown in Figure~\ref{fig:loopi} as a function of the $A^D$ and $m_Q/m_F$. In order to make contact with the effective parametrization adopted above, we have rescaled the obtained values by the bottom quark Yukawa coupling. All dimensionless coupling constants are set to $1/2$. Two-loop contributions of the size indicated by the fifth panel in Fig.~\ref{fig2} then require the scale of dark fermion to be close to that of the mediators, as well as sizable values of the trilinear coupling of the dark flavon to the messengers.
As for the three-loop contribution, the requirement $Y^{\D}_{13}=A^{\D} Y^{\D}_{12} Y^{\D}_{23}$ is always satisfied via a suitable choice of the $m_F/v_R$ ratio, regardless of the mass spectrum chosen for the particles in the dark sector.    
We also checked that introducing a further hierarchy between messengers and dark flavon does not appear to substantially modify the results of our analysis. However we choose not to rely on these results as the computational effort required to reach a comparable precision is too demanding.

\section{Novel phenomenological implications}
\label{sec:pheno}

To conclude, we briefly discuss the main phenomenological aspects of the proposed model. Although the signatures of this scenario at collider and low energy experiments match those of the original proposal\cite{Biswas:2017lyg,Biswas:2016jsh,Biswas:2015sha,Gabrielli:2014oya,Fabbrichesi:2017zsc,Gabrielli:2016cut,Barducci:2018rlx,Fabbrichesi:2017vma}, the presence of novel dark flavon interactions strongly modifies its possible cosmological implications. In particular, imposing the flavor symmetry in the messenger sector for both the up and down sectors, the scenario predicts a dark fermion mass spectrum that matches the SM fermion one upon an almost constant rescaling. 

The dark fermion mass spectrum is constrained by vacuum stability arguments and by the requirements of dark matter, which set the heaviest dark fermion mass (associated with the top quark) in the ballpark of 50-100 TeV \cite{Gabrielli:2013jka,Gabrielli:2016cut,Gabrielli:2016vbb}. As a result, the  masses of these particles span from a few GeV up to the TeV scale, with an even lighter spectrum component allowed for dark fermions associated with the SM lepton sector. We remark that whereas the $U(1)_D$ gauge invariance and the absence of corresponding charged currents forbid all dark fermion decays in the original model, the considered flavon interactions enable these processes in the present framework. This is an important difference in the phenomenological consequences of the two scenarios, which attributes novel and richer signatures to the model at hand. In fact, the tree-level exchange of a heavy dark flavon $F$ induces an effective 4-fermion interaction of the form
\bea
\mathcal{L}_{\rm eff}\sim \frac{\eta_A \eta^{\prime}_B }{\Lambda^2_F} [\bar{Q}^{A_i} Q^{A_{i-1}}][\bar{Q}^{B_k} Q^{B_{k-1}}]\, ,
\eea
where $\Lambda_F$ is a scale of the order of flavon mass, $i,k\in\{1,2,3\}$ and $A,B\in\{U,D\}$. Given the hierarchical mass spectrum of dark fermions, the dark fermions associated with the third SM generation can therefore decay into lighter states. If we assume that each up sector dark fermion is heavier than the corresponding down sector particle, as suggested by the SM Yukawa hierarchy, the $U(1)_D$ charges in Eq.(\ref{Qrel}) allow the following processes
\bea
Q^{\U_3} \to Q^{\U_2} +Q^{\U_2} +\bar{Q}^{\U_1}\, ,
\eea
followed by the $Q^{\U_2}$ decay
\bea
Q^{\U_2} \to Q^{\U_1} +Q^{\D_2} +\bar{Q}^{\D_1}\, ,
\eea
provided that the mass relation $m_{Q^{\U_2}} - m_{Q^{\U_1}} >2 m_{Q^{\D_2}} $ hold. Analogously, the decay $Q^{\D_3} \to Q^{\D_2} +Q^{\D_2} +\bar{Q}^{\D_1}$ could also be allowed in the dark sector, reducing the number of stable dark fermions associated with SM quarks to three: $Q^{\U_1},~Q^{\D_2},~Q^{\D_1}$. The new decay processes mediated by the dark flavon consequently restrict the mass of the heaviest stable dark fermion to the ${\cal O}(100~{\rm GeV})$ range, with important consequences for the viability of this particle as a dark matter candidate. In particular, these interactions reconcile the scenario with the theoretical upper bound on the dark matter mass that unitarity imposes for the thermal production of a relic abundance of elementary particles~\cite{Griest:1989wd}. Dedicated phenomenological analysis of the scenario can further constraint the dark fermion mass spectrum, for instance by analyzing the number of thermalized degrees of freedom throughout the different cosmological eras.

\section{Conclusions}

In the context of models for the radiative generation of SM Yukawa couplings, we have extended the framework originally proposed in \cite{Gabrielli:2013jka,Gabrielli:2016vbb} to accommodate a new mechanism for the origin of flavour mixing. The novelty of our work is in the presence of an additional scalar field, the dark flavon, which sources new loop diagrams that determine the off-diagonal Yukawa matrix elements. The resulting structure recovers the MFV ansatz in a natural way, owing to the additional loop suppression that  off-diagonal contributions have with respect to the diagonal ones. The same suppression ensures that the new terms bear a negligible impact on the flavour hierarchy regulated by the diagonal elements, generated here as detailed in the original framework \cite{Gabrielli:2013jka,Gabrielli:2016vbb}.  

In order to demonstrate the viability of the mechanism, we focused on the flavour hierarchy and mixing that characterize the SM quark sector. In more detail, starting on the flavour basis defined by the up-type quark mass eigenstates, we have introduced an effective parametrization of the down quark Yukawa coupling matrix to model the emerging loop structure. The effective parameter space was then constrained by using the current measurements of the mass spectrum of the down quarks and of the mixing angles contained in the CKM matrix. As a second step, we have evaluated the new two and three-loop diagrams induced by the dark flavon interactions, which source here the off-diagonal Yukawa interactions. By comparing the obtained expressions with the values of the effective parameters selected in the previous step, we have then shown that the current observations can be matched for perturbative values of the involved couplings on top of a mild mass hierarchy in the constituents of the dark sector.
Our findings have also highlighted the presence of two qualitatively different solutions to the quark flavour puzzle, related to the emergence of the down quark mass as a pure effect of the quark mixing. We argue that the mechanism can be straightforwardly extended to the lepton sector of the SM and to  possibly explain also the smallness of the neutrino mass scale. 

The phenomenology of the model scenarios at collider and low energy experiments largely overlaps with that of the original framework. The main difference is that the new dark flavon interactions allow the heaviest dark fermions to decay into the lightest states, reducing by net the maximal mass of the proposed dark matter candidates. As a consequence, the scenario is fully compatible with the theoretical upper bounds that hold for a relic abundance of elementary particles thermally produced during the evolution of the Universe. 
\section*{Acknowledgements} 
\label{sec:Acknowledgements}
The authors thank Matthias Steinhauser for providing the {\tt q2e} and {\tt EXP} packages and guidance in their use. The authors are supported by the European Union through the ERDF CoE grant TK133 and by the Estonian Research Council  through the grants PRG356 and MOBTT86. EG is affiliated to the Institute for Fundamental Physics of the Universe, Trieste, Italy.


\end{document}